\documentclass[twocolumn]{revtex4}
\usepackage{graphicx}
\DeclareMathAlphabet\mathbfcal{OMS}{cmsy}{b}{n}
\usepackage{enumerate}
\usepackage{amssymb}
\usepackage{amsmath}
\usepackage{amsfonts}
\usepackage{dcolumn}
\usepackage{bm}
\usepackage{framed}
\usepackage{braket}
\usepackage{color}
\usepackage{alphalph}
\usepackage{fancyhdr}
\usepackage{upgreek}

\begin{document}
\title{Electric directional steering of cathodoluminescence from graphene-based hydrid nanostructures}
\author{A. Ciattoni$^1$}
\email{alessandro.ciattoni@spin.cnr.it}
\author{C. Conti$^{2,3}$}
\author{A. Marini$^4$}
\affiliation{$^1$CNR-SPIN, c/o Dip.to di Scienze Fisiche e Chimiche, Via Vetoio, 67100 Coppito (L'Aquila), Italy}
\affiliation{$^2$CNR-ISC, Via dei Taurini 19, 00185, Rome, Italy}
\affiliation{$^3$Department of Physics, University Sapienza, Piazzale Aldo Moro 5, 00185, Rome, Italy}
\affiliation{$^4$Department of Physical and Chemical Sciences, University of L'Aquila, Via Vetoio, 67100 L'Aquila, Italy}
\date{\today}
\begin{abstract}
Controlling directional emission of nanophotonic radiation sources is fundamental to tailor radiation-matter interaction and to conceive highly efficient nanophotonic devices for on-chip wireless communication and information processing. Nanoantennas coupled to quantum emitters have proven to be very efficient radiation routers, while electrical control of unidirectional emission has been achieved through inelastic tunneling of electrons. Here we prove that the radiation emitted from the interaction of a high-energy electron beam with a graphene-nanoparticle composite has beaming directions which can be made to  continuously span the full circle even through small variations of the graphene Fermi energy. Emission directionality stems from the interference between the double cone shaped electron transition radiation and the nanoparticle dipolar diffraction radiation. Tunability is enabled since the interference is ruled by the nanoparticle dipole moment whose amplitude and phase are driven by the hybrid plasmonic resonances of the composite and the absolute phase of the graphene plasmonic polariton launched by the electron, respectively. The flexibility of our method provides a way to exploit graphene plasmon physics to conceive improved nanosources with ultrafast reconfigurable radiation patterns.

\end{abstract}

\maketitle
\count\footins = 1000

The ability of a nanophotonic radiation source to pump out streams of directional photons is a fundamental requirement enabling its efficient integration in a device or, more generally, in a nano-environment. Nanoantennas fed by nanoscale quantum emitters (e.g. fluorescent molecules, quantum dots, etc.) \cite{Novot,Giann,Biagi}
are probably the most interesting and widely investigated nanosources exhibiting directional emission. Basically the electromagnetic near field of the quantum emitter is converted by the nanoantenna into freely propagating optical radiation whose interference with the primary quantum emitter radiation provides overall directional emission. Radiation beaming has been achieved in a number of different designs as the Yagi-Uda \cite{Curto,Kosak,Ramez,Seeee}, planar \cite{Leeee}, leaky-wave \cite{Shega,Peter} and patch \cite{Belac,Yangg} nanoantennas. Single element nanoantennas with smaller footprints are provided by nanoparticles whose directional emission stems from the interference among their multipolar moments. High refractive index dielectric nanospheres and nanocylinders can be made to display beaming functionality \cite{Fuuuu,Curt2,Kuzne,Cihan} since their strong Mie multipolar resonances enable a suitable tailoring of multipoles interference. An analogous multipolar moments management can be achieved through plasmonic nanoparticles with asymmetric shape \cite{Hancu,Vercr} or hybrid composition \cite{Sheg2,Pelle,Rusak}.

All these schemes do not enable ultrafast active control of the emitted radiation pattern which is assigned by the geometry and composition of the nanosource with no input channel for electrical pulses. Electrically driven nanosources have been achieved by coupling a plasmonic nanoantenna with a metal–insulator–metal junction \cite{Parze} and electric control of the radiation spectrum has been observed \cite{Kernn,Buret,HeHee}. In such setups, the electrically biased junction triggers inelastic electron tunneling which generates light in turn exciting the nanoantenna. Since the bias voltage tunes the interference between the dipolar junction radiation and the nanoantenna multipolar moments, electric control of angular spread (directivity) of the emitted radiation has been obtained \cite{Gurun}. Electron-matter interaction provides alternative strategies for achieving directional emission of radiation. In a scanning tunneling microscope (STM), the interaction of low-energy electrons  with an asymmetric plasmonic nanoparticle both generates light by inelastic electron tunneling and selectively excites localized plasmonic modes of the nanoparticle, thus enabling radiation beaming \cite{LeMoa}. Analogously, the plasmonic modes of a nanoantenna can also be excited by a high-energy electron beam focused to subwavelength dimensions which also provides primary radiation through a cathodoluminescence process \cite{Coen1,Coen2,Coen3}. Remarkably, in both cases the emitted radiation pattern strongly depends on the position of the electron excitation (i.e. the STM tip and the electron beam axis, respectively) thus enabling control over the beaming direction.

\begin{figure*}  \label{Fig1}
\includegraphics[width=1\textwidth]{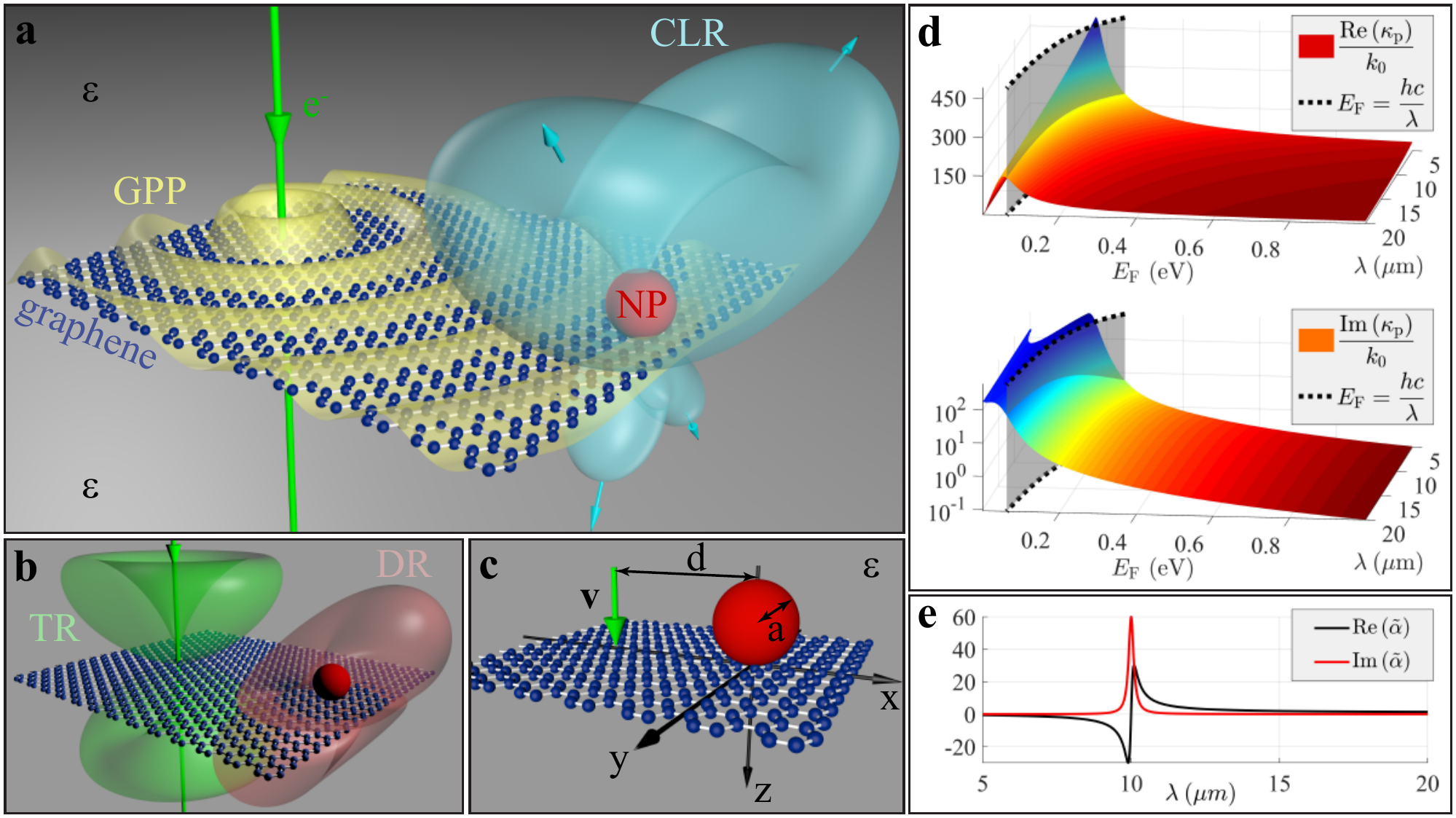}
\caption{\bf Cathodoluminescence directional emission from the interaction of a fast electron with the graphene-nanoparticle composite. (a) The electron ($e^-$) normally crossing the graphene sheet triggers the emission of TR (see panel (b)) and the excitation of a GPP which reaches the NP. The DR emitted by the NP (see panel (b)) interferes with the TR thus providing an overall CLR exibiting directions of maximal emission (arrows). (b) Schematic of the double cone shaped TR and the dipolar DR. (c) Geometrical parameters characterizing the setup. (d) Real and imaginary parts of the graphene plasmon wavenumber $\kappa_p$  normalized with the vacuum wave number $k_0$ as functions of the graphene Fermi energy $E_{\rm F}$ and the photon mid-infrared wavelength $\lambda$. $E_{\rm F}$ equates the photon energy $hc/\lambda$ along the dashed line which is the GPP excitation threshold. (e) Real and imaginary parts of the normalized NP polarizability $\tilde \alpha  = \alpha /\left( {4\pi \varepsilon _0 \varepsilon a^3 } \right)$ displaying the NP localized plasmonic resonance at $\lambda = 10 \, {\rm \mu m}$.}
\end{figure*}

Electrically driven nanoantennas considered so far mainly enable ultrafast modulation of the radiation emission and directivity tuning. On the other hand, the ability to electrically control the beaming direction over a broad solid angle is still missing to the best of our knowledge. In a typical setup the bias voltage basically affects the strength of the tunnel junction dipole, its direction being set almost entirely by the fixed junction geometry so that, since the nanoantenna geometry is fixed as well, the direction of the overall radiation emission very poorly depends on the electrical stimulation. In a paradigmatic nanosource scheme where a radiation emitter is coupled to a nanoantenna, ultrafast electric control of the beaming direction can only be achieved if the emitter and/or nanoantenna can be "electrically reoriented", i.e. if their multipoles can be fully driven by an electric pulse, an avenue not yet considered in literature.

In this paper we theoretically demonstrate that a graphene sheet evanescently coupled to a transparent semiconductor nanoparticle, when hit by a high-energy electron, yields cathodoluminescence radiation whose beaming directions can be made to continuously span a broad angular range by varying the graphene Fermi energy. A schematic of the considered graphene-nanoparticle composite is reported in Figs.1a. The composite is embedded in a transparent medium ($\varepsilon$) and, in the sub-Cherenkov regime we here focus on, the interaction between the fast electron ($e^-$) and the graphene sheet yields emission of transition radiation (TR, see Fig.1b) \cite{Misko,Zhang,Akbar} which is the primary radiation source in our setup. Electron-graphene interaction also triggers the emission of a mid-infrared graphene plasmon polariton (GPP) \cite{deAb1,Ochia,Wonhg,Linnn} which, after reaching the nanoparticle (NP), excites the hybrid plasmonic modes of the graphene-nanoparticle composite \cite{Monda,Ciatt}. As a consequence the diffraction radiation (DR, see Fig.1b) \cite{Potyl} emitted by the nanoparticle interfere with TR \cite{deAb2,Fiori,Sanno} thus providing an overall cathodoluminescence radiation (CLR) exhibiting directions of maximal emission, whose tuning is made possible by two distinct physical mechanisms. First, the hybrid plasmonic modes are sensitive to the graphene Fermi energy so that its variation, close to the hybrid resonances,  effectively produces a reorientation of the NP dipole moment and a consequent change of the maximal emission direction (TR being effectively independent on the Fermi energy). Second, the NP dipole moment is proportional to the GPP field so that the absolute plasmon phase, which is extremely sensitive to the graphene Fermi energy, directly drives the TR/DR interference and dramatically affects the directions of maximal emission to the point that even relatively small changes of the Fermi energy enable these directions to span the full circle around the electron trajectory.

{\bf RESULTS AND DISCUSSION}

{\bf Hybrid plasmonic modes excitation}. An electron of charge $-e$ moving with constant velocity $v$ in a medium of dielectric permittivity $\varepsilon$ produces a field whose spectral component at frequency $\omega$, assuming the electron velocity parallel to the $z$-axis, is
\begin{equation} \label{Ee}
 {\bf{E}}_\omega ^{\rm \left( e \right)}  = E_{\omega 0} \frac{e^{i \frac{\omega}{v} z } }{{\varepsilon \beta ^2 \gamma }}\left[ { - K_1 \left( {\frac{{\omega \rho }}{{v\gamma }}} \right){\bf{\hat e}}_\rho   + \frac{i}{\gamma }K_0 \left( {\frac{{\omega \rho }}{{v\gamma }}} \right){\bf{\hat e}}_z } \right]
\end{equation}
where  $\beta = v/c$, $\gamma  = 1/\sqrt{1 - \varepsilon \beta ^2}$ is the Lorentz contraction factor,  $E_{\omega 0}  = ek_0 Z_0 /4\pi ^2$ ($k_0 = \omega /c$ and $Z_0  = \sqrt {\mu _0 / \varepsilon _0}$ is the vacuum impedance), ${\bf{\hat e}}_\rho$ is the radial unit vector of cylindrical coordinates $(\rho,z)$ coaxial with the charge trajectory and $K_n$  are the modified Bessel function of the second kind. We here focus on the sub-Cherenkov regime $v < c/ \sqrt{ \varepsilon}$ where the exponential decay of ${\bf{E}}_\omega ^{\left( {\rm e} \right)}$ in the radial direction prevents emission of electromagnetic radiation. The field ${\bf{E}}_\omega ^{\rm \left( e \right)}$ comprises photons of frequency $\omega$ with normal wave vector $k_z = \omega /v$ (the only ones the electron is able to emit) and the distribution of their parallel wave vectors $k_{\parallel}$ (along the graphene sheet) is a Lorentzian of width $\Delta k_\parallel   = \omega /\left( {v\gamma } \right)$, due to the free photon dispersion relation $k_\parallel ^2  + k_ z^2  - k_0^2 \varepsilon = 0$.

If the electron normally crosses a graphene a sheet at $z=0$ in the absence of the NP (see Fig.1c), the overall field is 
\begin{equation} \label{Eeg}
{\bf{E}}_\omega ^{\left(\rm eg\right)}  = {\bf{E}}_\omega ^{\rm \left( e \right)}  + {\bf{E}}_\omega ^{\left( \rm {g} \right)}
\end{equation}
where ${\bf{E}}_\omega ^{\left( \rm {g} \right)}$ is the field produced by the graphene surface charge redistribution triggered by the moving electron. Such field is of a central importance in our analysis since both it accounts for the TR in the far field and it describes the GPP in the near field. As a matter of fact ${\bf{E}}_\omega ^{\left( \rm {g} \right)}$ is a source free field outside the graphene sheet and it is polarized in the radial plane (transverse magnetic) as ${\bf{E}}_\omega ^{\left( \rm {e} \right)}$. The distribution of its photon parallel wave vectors is the above Lorentzian of width $\Delta k_\parallel = k_0 /\left( {\beta \gamma} \right)$ ($ \equiv \omega/(v \gamma)$) with an additional Fresnel factor displaying a pole at the complex plasmon wavenumber $\kappa _{\rm p}  = k_0 \sqrt {\varepsilon  - \left( {\frac{{2\varepsilon }}{{Z_0 \sigma }}} \right)^2 }$ where $\sigma (k_{\parallel},\omega)$ is the graphene conductivity (see Supporting Information). In this paper we consider relativistic electrons with $\beta = 0.1$ ($2.69 \, {\rm KeV}$) since faster electrons produce TR so strong to prevent interference with DR (see below) and we set $\varepsilon = 2$ so that $\Delta k_\parallel \approx 10 \, k_0 $. As a consequence, the photons of ${\bf{E}}_\omega ^{\left( \rm {g} \right)}$ with $k_\parallel  < k_0 \sqrt 2$, which are able to reach the far field, are all efficiently excited and they set up the TR. In Fig.1d we plot the real and imaginary parts of the (normalized) plasmon wavenumber $\kappa _{\rm p}$ (as functions of the Fermi energy $E_{\rm F}$ and the mid-infrared wavelength $\lambda  = 2\pi c/\omega$) evaluated with the local model for the graphene conductivity $\sigma(\omega)$ (in the random phase approximation, see Supporting Information). If $E_{\rm F}$ is smaller than the photon energy $hc/\lambda$ (region at the left of the grey surface), graphene almost behaves as an absorbing dielectric, due to the onset of interband transition, and consequently the electron is not able to trigger graphene plasmonic resonances since ${\rm Re} \left( \kappa_{\rm p} \right) > 150 \: k_0 \gg \Delta k_{\parallel}$.  On the other hand, if $E_{\rm F}$ is greater that $hc/\lambda$, graphene behaves as a conductor with low absorption and accordingly ${\rm Re} \left( \kappa_{\rm p} \right)$ gets comparable with $\Delta k_{\parallel}$ and much greater than ${\rm Im} \left( \kappa_{\rm p} \right)$, this implying that a radially propagating GPP is launched by the electron crossing. Since GPP excitation is crucial for our purposes (see below) we hereafter focus on the regime $E_{\rm F} > hc/\lambda$ where, in addition, the local model for the graphene conductivity is fully adequate for the parallel wave vector range $\Delta k_\parallel$ of photons generated by the electron (see Supporting Information). 

The GPP field is also a very good approximation of the full electron-graphene field of Eq.(\ref{Eeg}) in the near field region surrounding the graphene sheet and far from the electron trajectory ($\rho \gg 1/\Delta k_\parallel$) where the modified Bessel function exponentially vanish (plasmon pole approximation). The GPP field turns out to be the contribution of the residue at the plasmon pole $\kappa_{\rm p}$ and it is given by
\begin{equation} \label{GPP}
{\bf{E}}_\omega ^{\left( {\rm eg} \right)}  = E_{\omega 0} \frac{{i\pi }}{{\varepsilon \beta }}\frac{{\kappa _{\rm p}^3 \left[ {H_1^{\left( 1 \right)} \left( {\kappa _{\rm p} \rho } \right){\bf{\hat e}}_\rho   + \frac{z}{{\left| z \right|}}H_0^{\left( 1 \right)} \left( {\kappa _{\rm p} \rho } \right){\bf{\hat e}}_z } \right]}}{{k_0 \left( {\kappa _{\rm p}^2  + \Delta k_\parallel ^2 } \right)}}e^{ - \kappa _{\rm p} \left| z \right|} 
\end{equation}
where $H_n^{(1)}$ are the Hankel functions of the first kind (see Supporting Information). Note that such GPP field is very sensitive to the plasmon wavenumber $\kappa_{\rm p}$ so that it is directly driven by electric bias.

\begin{figure*} \label{Fig2}
\center
\includegraphics[width=1\textwidth]{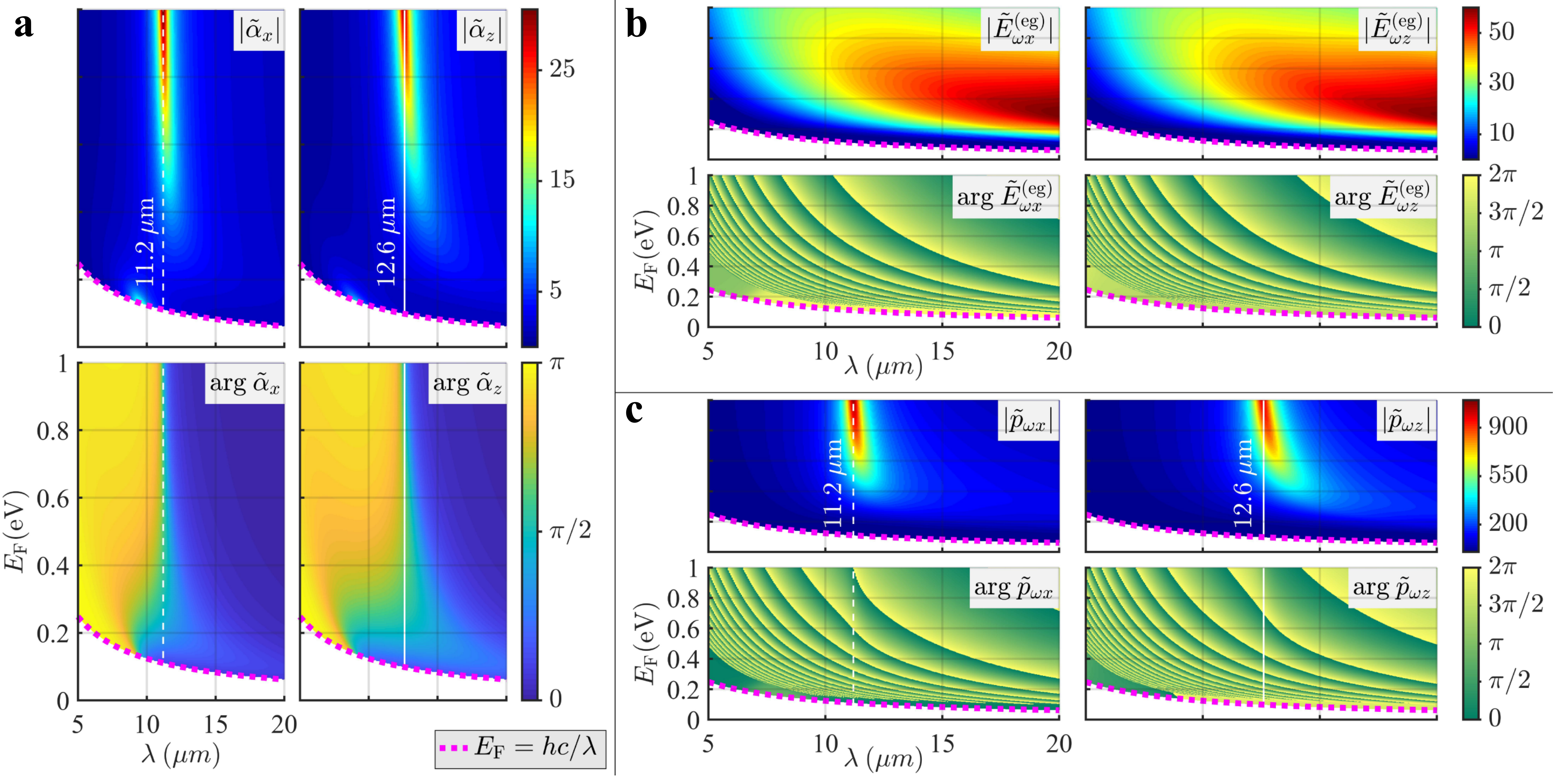}
\caption{\bf Excitation of hybrid plasmonic modes by electron crossing. We here consider the regime of Fermi energy greater than the photon energy and we set $\epsilon = 2$ for the hosting medium permittivity, $\beta = 0.1$ for the relativistic electron velocity and $d = 1000 \, {\rm nm}$ for the impact parameter. (a) Moduli and phases of the normalized effective polarizabilities $\tilde \alpha_j = \alpha_j /\left( {4\pi \varepsilon _0 \varepsilon a^3 } \right)$ of the NP evanescently coupled to graphene. At the hybrid plasmonic resonance wavelengths the polarizabilies phases have a $\pi$ jump. (b) Moduli and phases of the normalized components $\tilde E_{\omega j}^{(\rm eg)} = E_{\omega j}^{(\rm eg)} /E_{\omega 0}$ of the electron-graphene field at the NP center. (c) Moduli and phases of the normalized components $\tilde p_{\omega j} = \tilde \alpha_j \tilde E_{\omega j}^{(\rm eg)}$ of the dipole moment representing the NP excited by the electron. 
}
\end{figure*}

In the presence of the NP lying on the graphene sheet (see Fig.1c), the full electric field is
\begin{equation} \label{E}
{\bf{E}}_\omega   = {\bf{E}}_\omega ^{\left( {\rm eg} \right)}  + {\bf{E}}_\omega ^{\left( {\rm NP} \right)} 
\end{equation}
where ${\bf{E}}_\omega ^{\left( {\rm NP} \right)}$ is the field generated by the NP interacting with both the electron and the graphene sheet and it displays hybrid plasmonic resonances due to the NP-graphene evanescent coupling. The hybridization is effective when the NP supports localized plasmon modes in the mid-infrared so that we consider a transparent conducting oxide NP \cite{Liiii,Matsu,Wangg} of radius $a = 30 \: {\rm nm}$ with Drude dielectric pemittivity $\varepsilon_{\rm NP} (\omega) = 1 - \frac{\omega_{\rm p}^2}{\omega^2 + i \omega \Gamma}$, where we have chosen $\omega_p = 4.21 \cdot 10^{14} \, \rm{Hz}$, $\Gamma = 3.76 \cdot 10^{12} \, \rm{Hz}$ in such a way that $\varepsilon_{\rm NP} = -4 + 0.1i$ at $\lambda = 10 \, {\rm \mu m}$. Since the radius $a$ is much smaller than the mid-infrared wavelengths, in the non-retarded approximation we model the NP as a point dipole located at ${\bf r}_{\rm NP} = -a \hat{\bf e}_z$ (see Fig.1c) with dipole moment ${\bf{p}}_\omega   = \alpha {\bf{E}}_\omega ^{\left( {\rm ext} \right)}$  where ${\bf{E}}_\omega ^{\left( {\rm ext} \right)}$ is the field experienced by the dipole (without self-field) and $\alpha  = 4\pi \varepsilon _0 \varepsilon  a^3 \left( {\frac{{\varepsilon _{\rm NP}  - \varepsilon }}{{\varepsilon _{\rm NP}  + 2\varepsilon }}} \right)$ is the well-known sphere polarizability. In Fig.1e, we plot the real and imaginary parts of the (normalized) nanoparticle polarizability showing its plasmonic resonance at $\lambda = 10 \, {\rm \mu m}$ where ${\mathop{\rm Re}\nolimits} \left( {\varepsilon _{\rm NP}  + 2\varepsilon } \right) = 0$. The field generated by the dipole is
\begin{equation} \label{ENP}
{\bf{E}}_\omega ^{\left( {\rm NP} \right)}  = \left[\uptheta \left( { - z} \right)  {\left( {G^{\left( {\rm i} \right)}  + G^{\left( {\rm r} \right)} } \right)+ \uptheta \left( z \right)  {G^{\left( {\rm t} \right)} } } \right]{\bf{p}}_\omega  
\end{equation}
where the dyadics $G$ yield the fields that are incident on, reflected from and transmitted by the graphene sheet, whereas electron excitation and graphene coupling provide the dipole moment
\begin{equation} \label{dipole}
{\bf{p}}_\omega   = \left[ {\frac{ 1  }{ \frac{1}{\alpha } - G^{\left( {\rm r} \right)} }  }   {\bf{E}}_\omega ^{\left( {\rm eg} \right)} \right]_{{\bf{r}} = {\bf{r}}_{\rm NP} } 
\end{equation} 
(see Supporting Information). 

The hybridization of NP and grahene resonances is signaled by the vanishing of $\det \left[ \frac{1}{\alpha } - G^{\left( {\rm r} \right)} \left( {\bf{r}}_{\rm NP} \right) \right]$ in Eq.(\ref{dipole}) and it is efficient since the NP dipole is immersed in the evanescent resonant plasmon field. Since the electron trajectory has an impact parameter $d$ with respect to the NP center (see Fig.1c), the electron-graphene field ${\bf{E}}_\omega ^{\left( {\rm eg} \right)}$ at the dipole position has only $x$- and $z$- components and Eq.(\ref{dipole}) turns into $p_{\omega x}  = \alpha _x E_{\omega x}^{\left( {\rm eg} \right)}$, $p_{\omega y} = 0$ and $p_{\omega z}  = \alpha _z E_{\omega z}^{\left( {\rm eg} \right)}$ so that the NP induced dipole is elliptically polarized in the $xz$- plane. In Fig.2a we plot the moduli and phases of the (normalized) effective polarizabilities $\alpha_x$ and $\alpha_z$ in the regime of Fermi energies greater than the photon energy. The polarizabilities $\alpha_x$ and $\alpha_z$ are peaked and experience a phase variation of $\pi$ at $\lambda = 11.2 \, {\mu m}$ and $\lambda = 12.6 \, {\mu m}$, respectively, which are therefore the hybrid plasmonic resonance wavelengths. They are close to the bare NP resonance wavelength $\lambda = 10 \, {\mu m}$ and the splitting reveals efficient excitation of graphene plasmon polaritons (different from the above considered GPP excited by the electron) supporting the evanescent coupling. Note that the larger $E_{\rm F}$ the better the quality of the resonances since the excitation of graphene plasmon polaritons is facilitated at larger Fermi energies by the smaller values of both ${\rm Re} (\kappa_{\rm p})$ and ${\rm Im} (\kappa_{\rm p})$ (see Fig.1d). 

Since we aim at driving the NP by the GPP excited by the electron crossing, in our analysis we set the electron-NP impact parameter $d$ to be larger than $1/\Delta k_\parallel$ in the chosen mid-infrared spectral range so that the plasmon pole approximation holds. In Fig.2b we plot the moduli and phases of the (normalized) components $E_{\omega x}^{(\rm eg)}$ and $E_{\omega z}^{(\rm eg)}$ of the electron-graphene field  of Eq.(\ref{Eeg}) evaluated at the NP center for the impact parameter $d = 1000 \, {\rm nm}$ and we have checked that the overall field is very well approximated by the GPP field of Eq.(\ref{GPP}) with $\rho = d$ and $z=-a$. Note that the GPP field is effective in the regime $E_{\rm F} > hc/\lambda$ and its strength is larger at larger wavelength, in agreement with the above discussion. On the other hand the phases of its components turn out to be rapidly varying at the lower wavelengths and with level curves resembling those of ${\rm Re} (\kappa_{\rm p})$ (see Fig.1d). This can easily be grasped by noting that $\kappa_{\rm p} d \gg 1$ in the considered situation and hence the Hankel functions in Eq.(\ref{GPP}) can be replaced by their asymptotic expression thus yielding
\begin{equation}
{\bf{E}}_\omega ^{\left( {\rm eg} \right)}  = E_{\omega 0} \sqrt {\frac{{2\pi }}{{\kappa _{\rm p} d }}} \frac{{\kappa _{\rm p}^3 e^{ - \kappa _{\rm p} a} \left( {{\bf{\hat e}}_x  + i{\bf{\hat e}}_z } \right)}}{{\varepsilon \beta k_0 \left( {\kappa _{\rm p}^2  + \Delta k_\parallel ^2 } \right)}}e^{i\left( {\kappa _{\rm p} d - \frac{\pi }{4}} \right)} 
\end{equation}
which clearly shows that the GPP field responsible for the excitation of the hybrid plasmonic resonances has a global phase contribution equal to ${\rm Re} (\kappa_{\rm p} d)$. In Fig.2c we plot the moduli and the phases of the (normalized) components $p_{\omega x}$ and $p_{\omega z}$ of the NP induced dipole moment (see Eq.(\ref{dipole})) whose behavior evidently shows the features of both hybrid plasmonic modes and GPP field. In particular, the generally elliptical polarization of the dipole reduces to almost linear close to the resonance wavelengths and, most importantly, the dipole phase is extremely sensitive to $E_{\rm F}$ even off-resonance since it is driven by the GPP phase ${\rm Re} (\kappa_{\rm p} d)$.

\begin{figure} \label{Fig3}
\center
\includegraphics[width=0.5\textwidth]{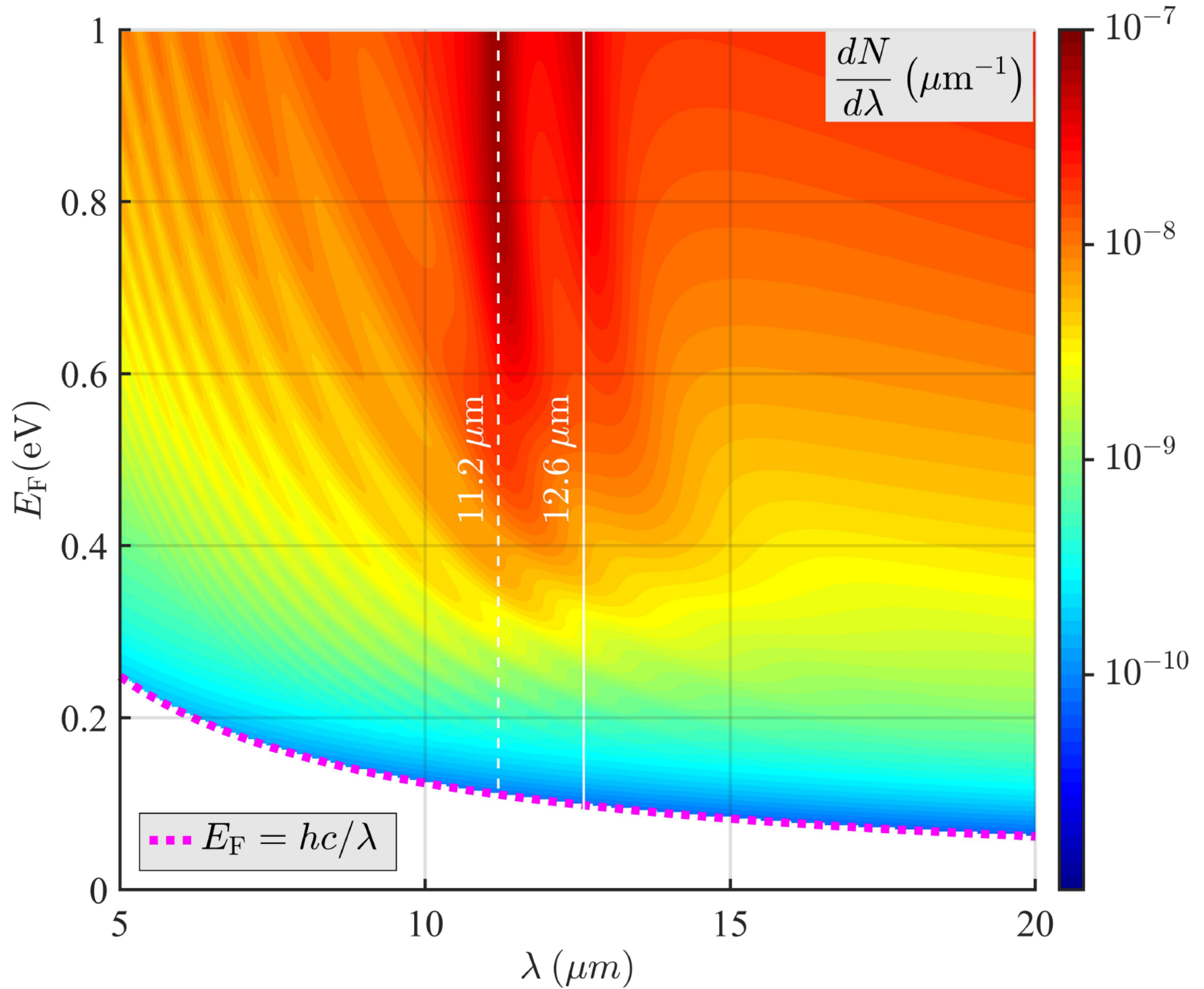}
\caption{\bf Electric tuning of CLR emission. $dN/d \lambda$ is the photon emission probability per unit of wavelength displaying peaks at the hybrid plasmon resonance wavelengths and oscillations due to the interference between TR and DR.}
\end{figure}

{\bf CLR emission}. The excitation of hybrid modes is produced by the electron-graphene-NP interactions in the near filed which are self-consistently mediated by the overall field ${\bf E}_\omega$ of Eq.(\ref{E}). Such near field coupling have a large impact on the CLR emission since this radiation is set up by the photons of the field ${\bf E}_\omega$ which survive in the far field. In the limit $k_0 r \rightarrow \infty$, the field gets the asymptotic expression ${\bf{E}}_\omega   = \frac{{e^{i\sqrt \varepsilon \left(k_0 r\right)} }}{{k_0 r}}\left[ {{\bf{f}}^{\left( {\rm g} \right)}  + {\bf{f}}^{\left( {\rm NP} \right)} } \right]$ with
\begin{eqnarray}  \label{ff}
 {\bf{f}}^{\left( {\rm g} \right)}  &=& E_{\omega 0} \left( {\frac{{Z_0 \sigma }}{{2\sqrt \varepsilon  }}} \right)\left( {\frac{{\beta \sin \theta \cos \theta }}{{\varepsilon \beta ^2 \cos ^2 \theta  - 1}}} \right){\bf{\hat e}}_\theta  , \nonumber \\ 
 {\bf{f}}^{\left( {\rm NP} \right)}  &=& \left( {{\bf{\hat e}}_\theta  {\bf{\hat e}}_\theta ^{\rm T}  + {\bf{\hat e}}_\varphi  {\bf{\hat e}}_\varphi ^{\rm T} } \right)\frac{{k_0^3 {\bf{p}}_\omega  }}{{4\pi \varepsilon _0 }}
\end{eqnarray}
where spherical coordinates $\left( r,\theta,\varphi \right)$ have been introduced together with their coordinate unit vectors ${\bf{\hat e}}_r$, ${\bf{\hat e}}_\theta$ and ${\bf{\hat e}}_\varphi$ and the dyadic notation $( {{\bf{ab}}^{\rm T} } ){\bf{c}} = \left( {{\bf{b}} \cdot {\bf{c}}} \right) {\bf{a}}$ has been used (see Supporting Information). Here ${\bf{f}}^{\left( {\rm g} \right)}$ and ${\bf{f}}^{\left( {\rm NP} \right)}$ are the far field amplitudes of the graphene and NP fields, ${\bf E}_\omega^{\rm (g)}$ and ${\bf E}_\omega^{\rm (NP)}$, respectively. Note that the bare electron field ${\bf E}_\omega^{\rm (e)}$ does not contribute to the CLR since we are considering the sub-Cherenkov regime. The amplitude ${\bf{f}}^{\left( {\rm g} \right)}$ describes the TR emitted by the graphene sheet upon interaction with the electron whereas ${\bf{f}}^{\left( {\rm NP} \right)}$ describes the  DR emitted by the NP upon excitation of hybrid modes (see Fig.1b). Since both amplitudes contribute to the far field and they are not orthogonal, the overall CLR intensity results from their interference and accordingly the photon emission probability is
\begin{equation} \label{emission}
\Gamma (\theta, \varphi) = \frac{{dN}}{{d\Omega d\lambda }}  = \frac{{\lambda \sqrt \varepsilon  }}{{\pi \hbar Z_0 }}\left[ {\left| {f_\theta ^{\left( {\rm g} \right)}  + f_\theta ^{\left( {\rm NP} \right)} } \right|^2  + \left| {f_\varphi ^{\left( {\rm NP} \right)} } \right|^2 } \right]
\end{equation}
which amounts to the number of photons emitted per incoming electron, per unit solid angle of emission $\Omega$, and per unit of photon wavelength $\lambda$ (see Supporting Information). In Fig.3 we plot the photon emission probability per unit of wavelength $dN/d \lambda$, obtained by integrating $\Gamma$ of Eq.(\ref{emission}) over the entire solid angle, clearly revealing two peaks of maximal emission occurring at the hybrid plasmonic resonance wavelengths $\lambda = 11.2 \, {\mu m}$ and $\lambda = 12.6 \, {\mu m}$. The sensitivity of the CLR emission to the GPP excitation is also particularly evident since $dN/d \lambda$ is globally larger in the upper region of the plane $(\lambda, E_{\rm F})$ where the plasmon wave vector $\kappa_p$ is almost real and with sufficiently small imaginary part. In addition $dN/ d\lambda$ displays oscillations whose period is larger at lower wavelengths and whose crests lie on curves resembling the level curves of ${\rm Re} (\kappa_{\rm p})$ (see Fig.1d). This behavior is a consequence of the interference between TR and DR since it results from the term $2{\mathop{\rm Re}\nolimits} ( {f_\theta ^{\left( {\rm g} \right)*} f_\theta ^{\left( {\rm NP} \right)} } )$ which is the only contribution in Eq.(\ref{emission}) depending on the GPP phase ${\rm Re} (\kappa_{\rm p} d)$ (carried by the factor $e^{i \kappa_{\rm p} d}$ in ${\bf{f}}^{\left( {\rm NP} \right)}$ due to ${\bf p}_\omega$). The electric tuning of CLR emission is therefore evident from Fig.3 and it arises from the tunability of the hybrid plasmon modes excitation enabled by GPP excitation.

\begin{figure*} \label{Fig4}
\center
\includegraphics[width=1\textwidth]{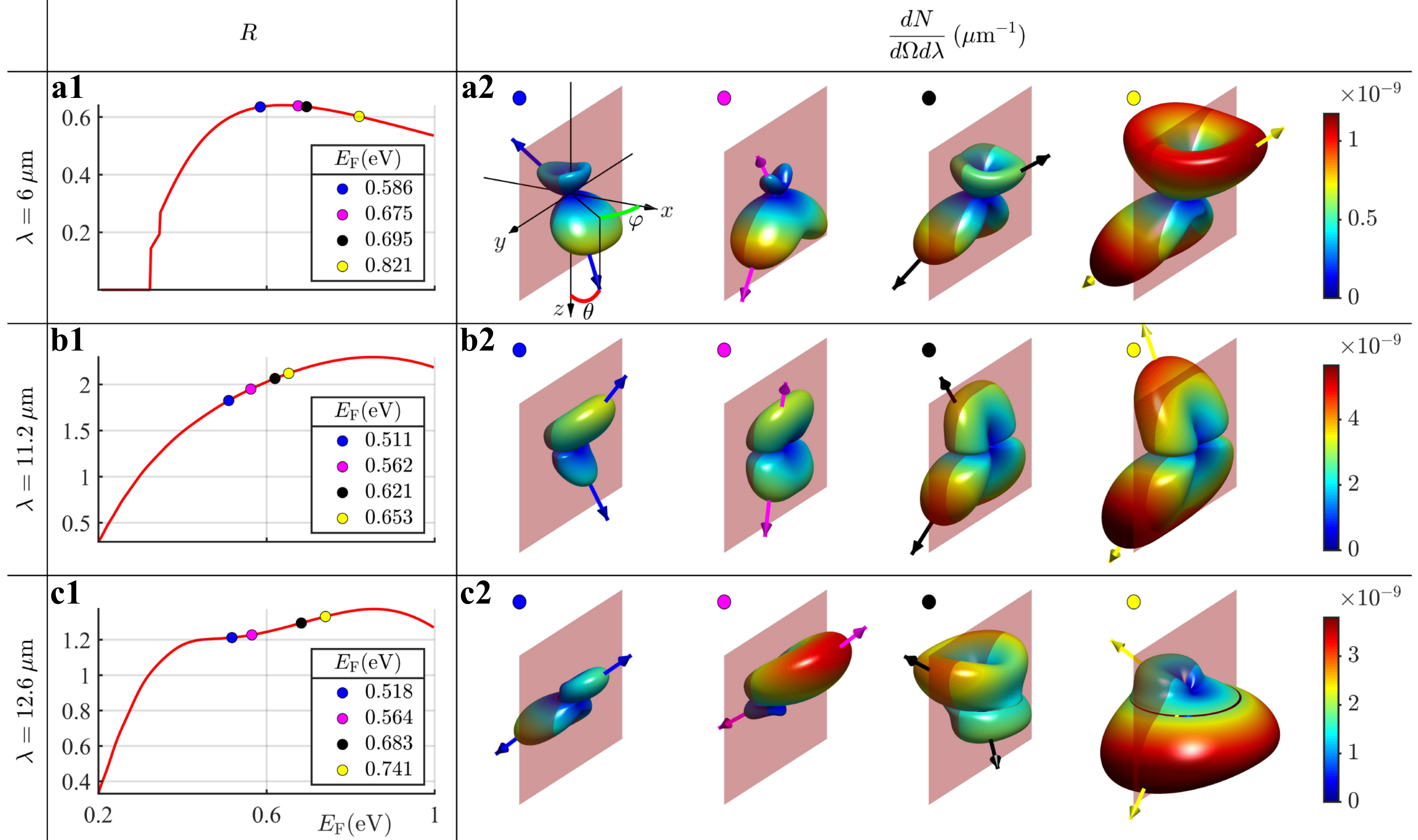}
\caption{\bf Electric tunability of CLR directional emission at an off-resonance wavelength (a) and at two on-resonance wavelengths (b) and (c). (a1,b1,c1) Ratio $R = | {f_\theta ^{\left( \rm p \right)} } |_{\max } / | {f_\theta ^{\left( \rm g \right)} } |_{\max }$ as function of the Fermi energy. (a2,b2,c2) Angle-resolved CLR emission patterns $\Gamma$ at the four specific values of $E_{\rm F}$ (blue, magenta, black and yellow circles) reported in the subplots of the first column. The arrows highlight the directions of maximal emission in the $y>0$ half-space whereas $(\theta,\varphi)$ (reported in the first of the a2 subplots) are the polar angles pertaining the direction of maximal emission in the quadrant $y>0$, $z>0$. 
}
\end{figure*}

{\bf Directional emission steering}. The interference between TR and DR supporting the above discussed oscillations of the CLR photon emission probability has a more far-reaching effect on the angular distribution of the emitted radiation. The  contributions of $| {{\bf{f}}^{\left( \rm g \right)} } |^2$,  $| {{\bf{f}}^{\left( \rm NP \right)} } |^2$ in Eq.(\ref{emission}) represent the angular distributions of TR and DR, respectively, whereas the term containing $2{\mathop{\rm Re}\nolimits} ( {f_\theta ^{\left( {\rm g} \right)*} f_\theta ^{\left( {\rm NP} \right)} } )$ accounts for their interference. TR is rotationally invariant around the $z$- axis and its double-cone profile (see Fig.1b) is only homothetically rescaled by varying the Fermi energy since the amplitude $f_{\theta}^{(\rm g)}$ is proportional to the graphene conductivity $\sigma$ (see the first of Eqs.(\ref{ff})). On the other hand DR has the usual electric dipole shape (see the second of Eq.(\ref{ff}) and Fig.1b) so that it is generally not rotational invariant around the $z$- axis, and it is ruled by the NP dipole ${\bf p}_\omega$ which, as detailed above, can be conveniently steered thorough the Fermi energy. Note that the electric control of DR is particularly effective close the plasmonic resonances and that it is not affected by GPP phase ${\rm Re} (\kappa_{\rm p} d)$. The interference term is even more interesting since it mixes the different geometrical features of the two amplitudes $f_\theta ^{\left( {\rm g} \right)}$ and $ f_\theta ^{\left( {\rm NP} \right)}$ and, most importantly, it is directly modulated by the GPP phase ${\rm Re} (\kappa_{\rm p} d)$. This is of central importance in our analysis since the GPP phase is the only quantity which is rapidly varying (with respect both $\lambda$ and $E_F$) on the overall chosen mid-infrared range. Since DR is not rotational invariant around the $z$- axis, as opposed to TR, their superposition displays directions of maximal emission which can be globally controlled through the Fermi energy. In other words CLR displays directional emission which can be controlled by electrically biasing the graphene sheet. Due to the role played by interference, such directional tunability is particularly effective when the strength of the TR and DR are comparable, i.e. when the ratio $
R = | {f_\theta ^{\left( \rm p \right)} } |_{\max } / | {f_\theta ^{\left( \rm g \right)} } |_{\max }$ is of the order of one, a condition which can be fulfilled by selecting  the electron energy (we here have chosen $\beta = 0.1$ in order to achieve this goal).

\begin{figure} \label{Fig5}
\center
\includegraphics[width=0.5\textwidth]{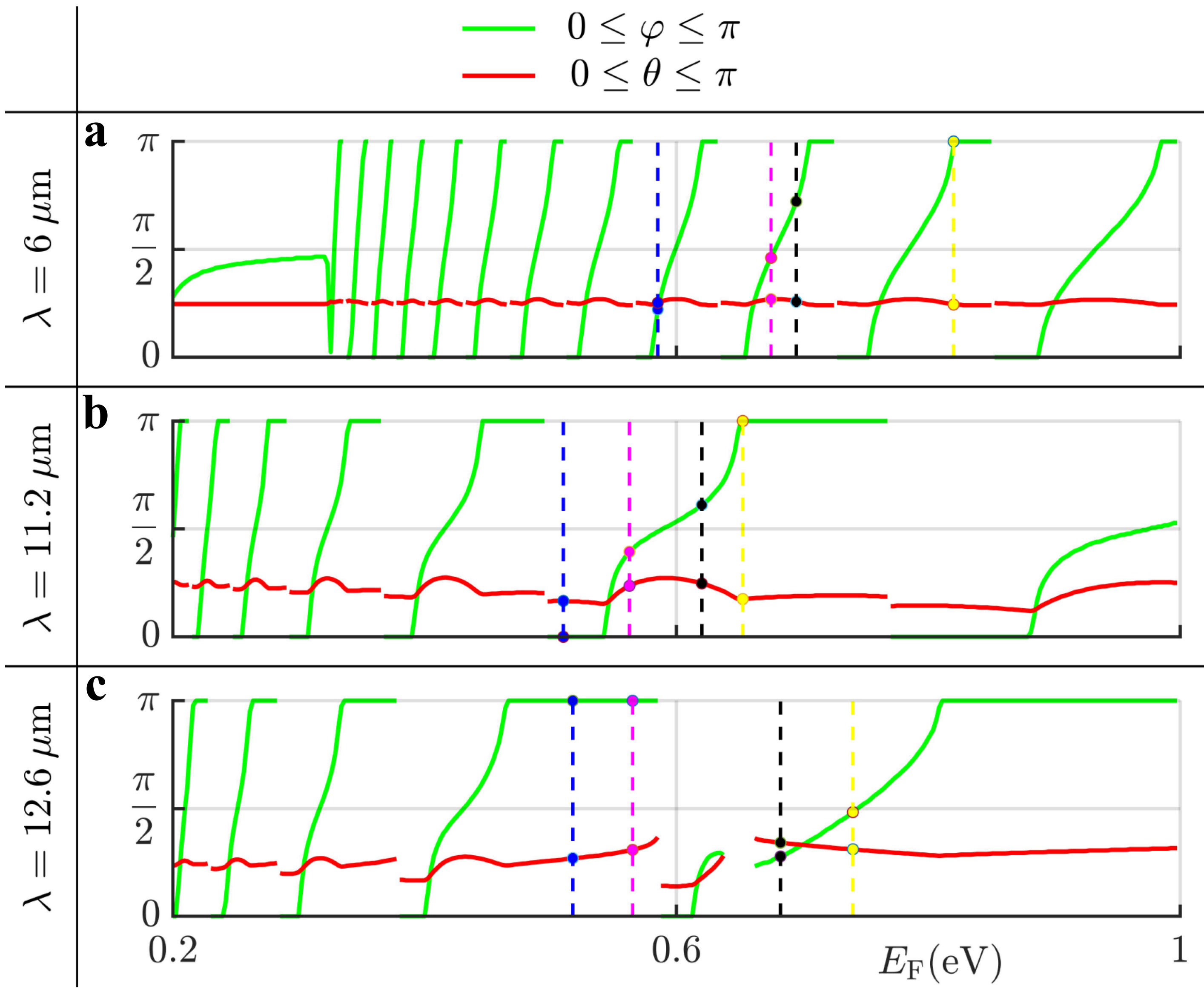}
\caption{\bf Electric tunability of CLR directional emission. Dependence of the polar angles $(\theta,\varphi)$ of the CLR maximal emission direction (see the first Fig.4a2 subplots) in the same three cases (a,b,c) considered in Fig.4. The vertical dashed lines and dots label the corresponding cases considered in Fig.4.}
\end{figure}

In Fig.4 we describe electric tunability of CLR directional emission at the off-resonance wavelength $\lambda = 6 \, {\rm \mu m}$ (a) and at the two on-resonance wavelengths $\lambda = 11.2 \, {\rm \mu m}$ (b) and $\lambda = 12.6 \, {\rm \mu m}$ (c). In each case we plot the ratio $R$ versus the Fermi energy (a1,b1,c1) and we depict the angle-resolved CLR emission pattern $\Gamma$ at four specific values of $E_{\rm F}$ (a2,b2,c2). Due to the setup invariance under reflection about the $xz$ plane, the emission patterns are invariant under $y \rightarrow -y$ and therefore we highlight with arrows only the directions of maximal emission in the $y>0$ half-space. We label with $(\theta,\varphi)$ the polar angles of the direction of maximal emission in the quadrant $y>0$, $z>0$ (see the first of the four subplots of Fig.4a2). In the off-resonance case (a) the ratio R is of the order of $0.5$ when the GPP is effectively excited ($E_{\rm F} > 0.4 \, {\rm eV}$) so that in this case the TR contribution is larger than the DR one. Accordingly the radiation patterns of Fig.4a2 have shapes qualitatively similar to the TR pattern but with relevant distortions due to DR which are different at different Fermi energies. Note that the direction of maximal emission in the $z >0$ half-space almost spans the full half-circle ($0<\varphi<\pi$) over the four considered situations. In both on-resonance case (b) and (c) the ratio R is larger than one in the GPP excitation regime so that the DR contribution is larger than the TR one. The case (b) corresponds to the excitation of the NP dipole which is almost linearly polarized along the $x$-axis (see Fig.2a) so that DR emission vanishes  along this axis as well as the TR one. Accordingly the CLR radiation pattern has effectively the shape of the $x$-polarized dipole radiation pattern but oppositely bent by TR in the half spaces $z>0$ and $z<0$ (since $f_\theta ^{\left( {\rm g} \right)} \left( {\theta } \right)
 =  -  f_\theta ^{\left( {\rm g} \right)} \left( {\pi  - \theta  } \right)$). Also in this case the directions of maximal emission almost span the full half-cirle. Analogously, the case (c) corresponds to the excitation of the NP dipole almost linearly polarized along the $z$-axis (see Fig.2a) so that the interference between TR and DR provides even more spectacular distorted profiles since the two separated radiation patterns vanish along orthogonal axes. Consequently the directions of maximal emission in the $z>0$ and $z<0$ half-space can be more efficiently tuned in an independent fashion.
 
In Fig.5 we plot the full dependence on the Fermi energy of the polar angles $(\theta,\varphi)$ pertaining the direction of maximal CLR emission in the quadrant $y>0$, $z>0$ (see the first of the subplots of Fig.4a2) in the same three cases (a,b,c) considered in Fig.4. The most  striking feature is that the azimuthal angle $\varphi$ spans the range $0<\varphi<\pi$ many times as $E_{\rm F}$ increases with a larger periodicity at lower wavelengths (see specifically subplot a of Fig.5). Such a versatile tunability is a consequence of the fact that the interference between TR and DR is mainly driven by the GPP phase, as shown above, so that the periodicity of the phase factor $e^{i \kappa_{\rm p} d}$ (on the $(\lambda,E_{\rm F})$ plane) produces the fast variations of the directions of maximal CLR emission. This shows that even relatively small variations of the Fermi energy can provide the tuning of the directional emission over the full circle around the $z$-axis. In addition, in the on-resonance cases (b) and (c) the further dependence of the dipole strength $|{\bf p}_\omega|$ on $E_{\rm F}$ additionally provides more complex directional tunability features across the emerge of the resonances (say $E_{\rm F} > 0.4 \, {\rm eV}$).

In the Supporting Movies S1, S2 and S3 we provide the full dependence of the angle-resolved CLR emission patterns on the Fermi energy corresponding to three cases (a,b,c) of Figs.4 and 5, respectively (the CLR emission patterns of Fig.4 are selected snapshots of the corresponding Supporting Movies). In the Movies we also sketch the evolution of the separated TR and DR emission patterns as $E_{\rm F}$ increases together with the evolution of the dipole polarization ellipse ${\bf{p}}\left( t \right) = {\mathop{\rm Re}\nolimits} \left( {{\bf{p}}_\omega  e^{ - i\omega t} } \right)$ and of the moduli and phases of $p_{\omega x}$ and $p_{\omega z}$. The Movies emphasize more clearly that the variation rate with $E_{\rm F}$ of the TR and DR emission patterns coincides with the variation rate of the moduli $|p_{\omega x}|$ and $|p_{\omega z}|$ whereas the overall CLR emission patters follows the rapid variations of $\arg p_{\omega x}$ and $\arg p_{\omega z}$ due to the above discussed GPP phase driven interference mechanism.

{\bf CONCLUSION}

In conclusion we have shown that the interaction of an high energy electron with a graphene-nanoparticle composite is accompanied by directional emission of CLR whose directionality can be feasibly steered through electrically biasing the graphene sheet. We have shown that this is basically an interferometric effect where the TR and DR interfering components display very different dependences on the graphene Fermi energy. While TR is shape-invariant, the DR angular profile crucially depends on the Fermi energy through the excitation of hybrid plamonic modes triggered by electron crossing. Specifically DR is very sensitive to the phase of the GPP launched by the electron  which in turn interferometrically drives the direction of maximal CLR emission. In the nanoantennas language, TR plays the role of the primary radiation source whereas the DR is the radiation outcoupled by the NP nanoantenna which can be effectively "reoriented" and "phase-modulated" by the GPP. Due to its feasibility and flexibility, our method can easily be extended to more complex  setups (more than one nanoparticle, inclination of the electron trajectory, etc.) able to provide in principle a full control of the radiation directionality. We believe that our results could open novel avenues to conceive nanophotonic devices where the ultra-fast directional routing of signals is achieved at the deep subwavelength scale.

{\bf ACKNOWLEDGEMENTS} 

A.C and C.C. acknowledge PRIN 2017 PELM (grant number 20177PSCKT). A.M. acknowledges support from the Rita Levi Montalcini Fellowship (grant number PGR15PCCQ5) funded by the Italian Ministry of Education, Universities and Research (MIUR).

%
%la pittura che segue è quindi
%1) la TR è varia solo in ampiezza nel piano dei parametri 
%2) la DR è ha i picchi di risonanza e la fase che varia con la fase del plasmone. (Tuning della fase della nanoantenna, RIORIENTAZIONE ELETTRICA).


\begin{thebibliography}{10}
\newcommand{\enquote}[1]{``#1''}
\bibitem{Novot} Novotny, L. \& van Hulst, N. F. 
                Antennas for light. 
                {\it Nature Photon.} {\bf 5}, 83-90 (2011).
\bibitem{Giann} Giannini, V., Fernandez-Domínguez, A. I., Heck, S. C. \& Maier, S. A. 
                Plasmonic Nanoantennas: Fundamentals and Their Use in Controlling the Radiative 
                Properties of Nanoemitters. 
                {\it Chem. Rev.} {\bf 111}, 3888–3912 (2011).
\bibitem{Biagi} Biagioni, P., Huang, J.-S., \& Hecht, B.
                Nanoantennas for visible and infrared radiation.
                {\it Rep. Prog. Phys.} {\bf 75}, 024402 (2012)
\bibitem{Curto} Curto, A. G., Volpe, G., Taminiau, T. H., Kreuzer, M. P., Quidant, R. \& van Hulst, N. F. 
                Unidirectional Emission of a Quantum Dot Coupled to a Nanoantenna.
                {\it Science} {\bf 329}, 930-933 (2010).     
\bibitem{Kosak} Kosako, T., Kadoya, Y. \& Hofmann, H. F. 
                Directional control of light by a nano-optical Yagi–Uda antenna.
                {\it Nature Photon.} {\bf 4}, 312-315 (2010).
\bibitem{Ramez} Ramezani, M., Casadei, A., Grzela, G., Matteini, F., Tutuncuoglu, G., Ruffer, D., Morral, A. F. \& Rivas, J. G. 
                Hybrid Semiconductor Nanowire–Metallic Yagi-Uda Antennas.
                {\it Nano Lett.} {\bf 15}, 4889-4895 (2015).
\bibitem{Seeee} See, K.-M., Lin, F.-C., Chen, T.-Y., Huang, Y.-X., Huang, C.-H., Yesilyurt, A. T. M., \& Huang, J.-S.                
                Photoluminescence-Driven Broadband Transmitting Directional Optical Nanoantennas.
                {\it Nano Lett.} {\bf 18}, 6002-6008 (2018).
\bibitem{Leeee} Lee, K. G., Chen, X. W., Eghlidi, H., Kukura, P., Lettow, R., Renn, A., Sandoghdar, V. \& Gotzinger, S.
                A planar dielectric antenna for directional single-photon emission and near-unity collection efficiency. 
                {\it Nature Photon.} {\bf 5}, 166-169 (2011).
\bibitem{Shega} Shegai, T., Miljkovic, V. D., Bao, K., Xu, H., Nordlander, P., Johansson, P. \& Kall, M. 
                Unidirectional Broadband Light Emission from Supported Plasmonic Nanowires.
                {\it Nano Lett.} {\bf 11}, 706–711 (2011). 
\bibitem{Peter} Peter, M., Hildebrandt, A., Schlickriede, C., Gharib, K., Zentgraf, T., Forstner, J. \& Linden, S.
                Directional Emission from Dielectric Leaky-Wave Nanoantennas.
                {\it Nano Lett.} {\bf 17}, 4178-4183 (2017).
\bibitem{Belac} Belacel, C., Habert, B., Bigourdan, F., Marquier, F., Hugonin, J.-P., Michaelis de Vasconcellos, S.,
 Lafosse, X., Coolen, L., Schwob, C., Javaux, C., Dubertret, B., Greffet, J.-J., Senellart, P., \& Maitre, A. 
                Controlling Spontaneous Emission with Plasmonic Optical Patch Antennas.               
                {\it Nano Lett.} {\bf 13}, 1516–1521 (2013).
\bibitem{Yangg} Yang, Y., Li, Q. \& Qiu, M.     
                Controlling the angular radiation of single emitters using dielectric patch nanoantennas.         
                {\it App. Phys. Lett.} {\bf 107}, 031109 (2015).
\bibitem{Fuuuu} Fu, Y. H., Kuznetsov, A. I., Miroshnichenko, A. E., Yu,  Y. F. \& Luk’yanchuk, B.
                Directional visible light scattering by silicon nanoparticles
                {\it Nature Commun.} {\bf 4}, 1527 (2013).
\bibitem{Curt2} Curto, A. G., Taminiau, T. H., Volpe, G., Kreuzer, M. P., Quidant, R. \& van Hulst, N. F. 
                Multipolar radiation of quantum emitters with nanowire optical antennas.
                {\it Nature Commun.} {\bf 4}, 1750 (2013).
\bibitem{Kuzne} Kuznetsov, A. I., Miroshnichenko, A. E., Brongersma, M. L., Kivshar, Y. S. \& Luk’yanchuk, B.                 
                Optically resonant dielectric nanostructures.
                {\it Science} {\bf 354}, aag2472 (2016).
\bibitem{Cihan} Cihan, A. F., Curto, A. G., Raza, S., Kik, P. G. \& Brongersma, M. L.      
                Silicon Mie resonators for highly directional light emission from monolayer MoS2.
                {\it Nature Photon.} {\bf 12}, 284–290 (2018).
\bibitem{Hancu} Hancu, I. M., Curto, A. G., Castro-Lopez, M., Kuttge, M. \& van Hulst, N. F. 
                Multipolar Interference for Directed Light Emission.
                {\it Nano Lett.} {\bf 14}, 166-171 (2014).
\bibitem{Vercr} Vercruysse, D., Zheng, X., Sonnefraud, Y., Verellen, N., Di Martino, G., Lagae, L., Vandenbosch, G. A. E., Moshchalkov, V. V., Maier, S. A. \& Van Dorpe, P. 
                Directional Fluorescence Emission by Individual V‑Antennas Explained by Mode Expansion.
                {\it Acs Nano} {\bf 8}, 8232–8241 (2014).

\bibitem{Sheg2} Shegai, T., Chen, S., Miljkovic, V. D., Zengin1, G., Johansson, P. \& Kall, M. 
                A bimetallic nanoantenna for directional colour routing.
                {\it Nature Commun.} {\bf 2}, 481 (2011).               
\bibitem{Pelle} Pellegrini, G., Mazzoldi, P. \& Mattei, G. 
                Asymmetric Plasmonic Nanoshells as Subwavelength Directional Nanoantennas and Color Nanorouters: A Multipole Interference Approach.
                {\it J. Phys. Chem. C} {\bf 116}, 21536-21546 (2012).                
\bibitem{Rusak} Rusak, E., Staude, I., Decker, M., Sautter, J., Miroshnichenko, A. E., Powell, D. A., Neshev, D. N. \& Kivshar, Y. S.               
                Hybrid nanoantennas for directional emission enhancement.
                {\it App. Phys. Lett.} {\bf 105}, 221109 (2014).
\bibitem{Parze} Parzefall, M. \& Novotny, L.
                Optical antennas driven by quantum tunneling: a key issues review.
                {\it Rep. Prog. Phys.} {\bf 82} 112401 (2019).
\bibitem{Kernn} Kern, J., Kullock, R., Prangsma, J., Emmerling, M., Kamp, M. \& Hecht,B. 
                Electrically driven optical antennas.
                {\it Nature Photon.} {\bf 9}, 582–586 (2015).                
\bibitem{Buret} Buret, M., Uskov, A. V., Dellinger, J., Cazier, N., Mennemanteuil, M.-M., Berthelot, J., Smetanin, I. V., Protsenko, I. E., Colas-des-Francs,  G. \& Bouhelier, A.   
                Spontaneous Hot-Electron Light Emission from Electron-Fed Optical Antennas.
                {\it Nano Lett.} {\bf 15}, 5811-5818 (2015).
\bibitem{HeHee} He, X., Tang, J., Hu, H., Shi, J., Guan, Z., Zhang, S. \& Xu, H.
                Electrically Driven Optical Antennas Based on Template Dielectrophoretic Trapping.
                {\it ACS Nano} {\bf 13}, 14041-14047 (2019).
\bibitem{Gurun} Gurunarayanan, S. P., Verellen, N., Zharinov, V. S., Shirley, F. J., Moshchalkov, V. V., Heyns, M., Van de Vondel, J., Radu, I. P. \& Van Dorpe, P.                 
                Electrically Driven Unidirectional Optical Nanoantennas.
                {\it Nano Lett.} {\bf 17}, 7433-7439 (2017).
\bibitem{LeMoa} Le Moal, E., Marguet, S., Rogez, B., Mukherjee, S., Dos Santos, P., Boer-Duchemin, E., Comtet, G. \& Dujardin, G. 
                An Electrically Excited Nanoscale Light Source with Active Angular Control of the Emitted Light.
                {\it Nano Lett.} {\bf 13}, 4198–4205 (2013).
\bibitem{Coen1} Coenen, T., Vesseur, E. J. R., Polman, A. \& Koenderink, A. F.
                Directional Emission from Plasmonic Yagi-Uda Antennas Probed by Angle-Resolved Cathodoluminescence Spectroscopy.
                {\it Nano Lett.} {\bf 11}, 3779-3784 (2011).                
\bibitem{Coen2} Coenen, T., Vesseur, E. J. R. \& Polman, A.
                Deep Subwavelength Spatial Characterization of Angular Emission from Single-Crystal Au Plasmonic Ridge Nanoantennas.                
                {\it ACS Nano} {\bf 6}, 1742-1750 (2012).
\bibitem{Coen3} Coenen, T., Arango, F. B., Koenderink, A. F. \& Polman, A.
                Directional emission from a single plasmonic scatterer.
                {\it Nature Commun.} {\bf 5}, 3250 (2014).       
\bibitem{Misko} Mišković, Z. L., Segui, S., Gervasoni, J. L. \& Arista, N. R.
                Energy losses and transition radiation produced by the interaction of charged particles with a graphene sheet.
                {\it Phys. Rev B} {\bf 94}, 125414 (2016).
\bibitem{Zhang} Zhang, K.-C., Chen, X.-X., Sheng, C.-J., Ooi, K. J. A., Ang, L. K. \& Yuan, X.-S.   
                Transition radiation from graphene plasmons by a bunch beam in the terahertz regime.
                {\it Opt. Express} {\bf 25}, 20477 (2017).
\bibitem{Akbar} Akbari, K., Segui, S., Mišković, Z. L., Gervasoni, J. L. \& Arista, N. R.    
                Energy losses and transition radiation in graphene traversed by a fast charged particle under oblique incidence.
                {\it Phys. Rev B} {\bf 98}, 195410 (2018).           
\bibitem{deAb1} Garcìa de Abajo, F. J. 
                Multiple Excitation of Confined Graphene Plasmons by Single Free Electrons.
                {\it Acs Nano} {\bf 7}, 11409–11419 (2013).
\bibitem{Ochia} Ochiai, T. 
                Efficiency and Angular Distribution of Graphene-Plasmon Excitation by Electron Beam.
                {\it J. Phys. Soc. Jpn.} {\bf 83}, 054705 (2014).                
\bibitem{Wonhg} Wong, L. J., Kaminer, I., Ilic, O., Joannopoulos, J. D. \& Soljačić, M. 
                Towards graphene plasmon-based free-electron infrared to X-ray sources.
                {\it Nature Photonics} {\bf 10}, 46–52 (2016).    
\bibitem{Linnn} Lin, X., Kaminer, I., Shi, X., Gao, F., Yang, Z., Gao, Z., Buljan, H., Joannopoulos, J. D., Soljačić, M., Chen, H. \& Zhang, B. 
                Splashing transients of 2D plasmons launched by swift electrons
                {\it Sci. Adv.} {\bf 3}, e1601192 (2017).              
\bibitem{Monda} Mondal, A. \& Jana, N. R.
                {\it Rev. Nanosci. Nanotechnol.} {\bf 3}, 177 (2014).                
\bibitem{Ciatt} Ciattoni, A., Conti, C., Zayats, A. V. \& Marini, A. 
                Electric Control of Spin-Orbit Coupling in Graphene-Based Nanostructures with Broken Rotational Symmetry. 
                {\it Laser Photonics Rev.} {\bf 14}, 2000214 (2020).         
\bibitem{Potyl} Potylitsyn, A. P., Ryazanov, M. I., Strikhanov, M. N. \& Tishchenko, A. A.   
                Radiation from Relativistic Particles.
                {\it Springer-Verlag} GmbH Berlin Heidelberg, 2011.
\bibitem{deAb2} Garcìa de Abajo, F. J. 
                Optical excitations in electron microscopy.
                {\it Rev. Mod. Phys.} {\bf 82}, 211-275 (2010).             
\bibitem{Fiori} Fiorito, R. B., Shkvarunets, A. G., Watanabe, T., Yakimenko, V. \& Snyder, D. 
                Interference of diffraction and transition radiation and its application as a beam divergence diagnostic.
                {\it Phys. Rev. ST AB} {\bf 9}, 052802 (2006).        
\bibitem{Sanno} Sannomiya, T., Konecna, A., Matsukata, T., Thollar, Z., Okamoto, T., Garcìa de Abajo, F. J. \&  Yamamoto, N.  
                Cathodoluminescence Phase Extraction of the Coupling between Nanoparticles and Surface Plasmon Polaritons.
                {\it Nano Lett.} {\bf 20}, 592-598 (2020).   
\bibitem{Liiii} Li, S. Q., Guo, P., Zhang, L., Zhou, W., Odom, T. W., Seideman, T., Ketterson,  J. B. \& Chang, R. P. H. 
                Infrared Plasmonics with Indium Tin-Oxide Nanorod Arrays.
                {\it Acs Nano} {\bf 5}, 9161–9170 (2011).
\bibitem{Matsu} Matsui, H. \& Tabata, H.
                Oxide semiconductor nanoparticles for infrared plasmonic applications.
                in {\it Science and Applications of Tailored Nanostructures}, One Central Press, Altrincham, (2012), pp. 68–86.       
\bibitem{Wangg} Wang, Y., Overvig, A. C., Shrestha, S., Zhang, R., Wang, R., Yu, N. \& Dal Negro, L. 
                Tunability of indium tin oxide materials for mid-infrared plasmonics applications.
                {\it Opt. Mat. Express} {\bf 7}, 2727-2739 (2017).
\end{thebibliography}
\end{document}

\title{Electric directional steering of cathodoluminescence from graphene-based hydrid nanostructures: Supporting  Information}
\author{A. Ciattoni$^1$}
\email{alessandro.ciattoni@spin.cnr.it}
\author{C. Conti$^{2,3}$}
\author{A. Marini$^4$}
\affiliation{$^1$CNR-SPIN, c/o Dip.to di Scienze Fisiche e Chimiche, Via Vetoio, 67100 Coppito (L'Aquila), Italy}
\affiliation{$^2$CNR-ISC, Via dei Taurini 19, 00185, Rome, Italy}
\affiliation{$^3$Department of Physics, University Sapienza, Piazzale Aldo Moro 5, 00185, Rome, Italy}
\affiliation{$^4$Department of Physical and Chemical Sciences, University of L'Aquila, Via Vetoio, 67100 L'Aquila, Italy}
\date{\today}
\begin{abstract}
This document provides supporting information to "Electric directional steering of cathodoluminescence from graphene-based hydrid nanostructures". We present here the analytical description of the interaction of the high energy electron with the graphene-nanoparticle composite. After discussing the local model for the graphene conductivity and its limitations, we provide a detailed description of the electron-graphene interaction (together with the plasmon pole approximation), the excitation of the graphene-nanoparticle hybrid plasmonic modes and the cathodoluminescence photon emission probability.
\end{abstract}

\maketitle

\count\footins = 1000

\section{Infrared graphene local optical response}
At infrared frequencies the optical response of graphene is dominated by the conical band structure ${\cal E} = \pm v_{\rm F} |{\bf p}_\parallel|$ around the two Dirac points of the first Brillouin zone, where $v_{\rm F} \approx 9 \cdot 10^5$ m$/$s is the Fermi velocity and ${\cal E},{\bf p}_\parallel$ are the electron energy and momentum, respectively. In the present investigation we mainly focus on the infrared range of wavelengths $5 \: \mu m < \lambda < 20 \: \mu m$. While in undoped graphene the Fermi energy lies at the Dirac points, injection of charge carriers through electrical gating \cite{Chen2011} or chemical doping \cite{Liu2011} efficiently shifts the Fermi level up to $E_{\rm F} \approx 1$ eV owing to the conical dispersion and the 2D electron confinement. The response of graphene to photons of energy $\hbar \omega$ and in-plane momentum $\hbar {\bf k}_{\parallel}$ is described by the surface conductivity $\sigma({\bf k}_{\parallel},\omega)$  which is generally affected both by intraband and interband electron dynamics. The dependence of $\sigma$ on ${\bf k}_{\parallel}$ physically arises from electron-hole pairs excitation and it generally yields unwanted absorption (Landau damping) and nonlocal effects. However, if 
\begin{equation} \label{locality}
\frac{{ {k_ \parallel  } }}{{k_{\rm F} }} < \frac{{\hbar \omega }}{{E_{\rm F} }} < 2 - \frac{{ {k_ \parallel  } }}{{k_{\rm F} }}
\end{equation}
where $k_{\rm F}  = E_{\rm F} /\hbar v_{\rm F}$ is the Fermi wave number, the photon momentum is too small to trigger intraband transitions and interband transitions are forbidden by the Pauli exlusion principle \cite{Hwang}. Once interacting with photons satisfying Eq.(\ref{locality}), nonlocal effects can be neglected and graphene displays a marked metal-like behavior with long relaxation time $\tau = \mu E_{\rm F}/ev_{\rm F}^2$, where $\mu$ is the electron mobility, which conversely to noble metals can reach the picosecond time scale at moderate doping and purity (affecting electron mobility) \cite{JavierACSPhot}. In such local regime, random phase approximation (RPA) provides for the graphene conductivity the integral expression
\begin{equation} \label{sigma}
\sigma(\omega) = \frac{-ie^2}{\pi\hbar^2(\omega+i/\tau)}\int_{-\infty}^{+\infty}d{\cal E} \left\{|{\cal E}| \frac{ \partial f_{\cal E} }{\partial {\cal E}} +    \frac{ {\rm sign} \left(\cal E\right) }{1 - 4 {\cal E}^2/[\hbar(\omega+i/\tau)]^2} f_{\cal E} \right\},
\end{equation}
where $f_{\cal E} = \{ {\rm exp}[({\cal E} - E_{\rm F})/k_{\rm B} T] +1 \}^{-1}$ is the Fermi function ($k_{\rm B}$ is the Boltzmann constant and $T$ is the temperature). In our analysis, we focus on photon-graphene interactions satysfing Eq.(\ref{locality}) and accordingly we model graphene surface conductivity by means of Eq.(\ref{sigma}).

It is worth noting that, in the regime where the Fermi energy is greater than the photon energy ($E_{\rm F}  > \hbar \omega$), Eq.(\ref{locality}) can be casted as 
\begin{equation} \label{locality2}
k_\parallel   < \left( {\frac{c}{{v_F }}} \right)k_0  \simeq 333\,k_0,
\end{equation}
where $k_0 = \omega/c$ (vacuum wavenumber), which specifies the  wavevector range of those photons not triggering  nonlocal effects at frequency $\omega$.

\begin{figure}[!t] \label{Fig1}
\center
\includegraphics[width=0.6\textwidth]{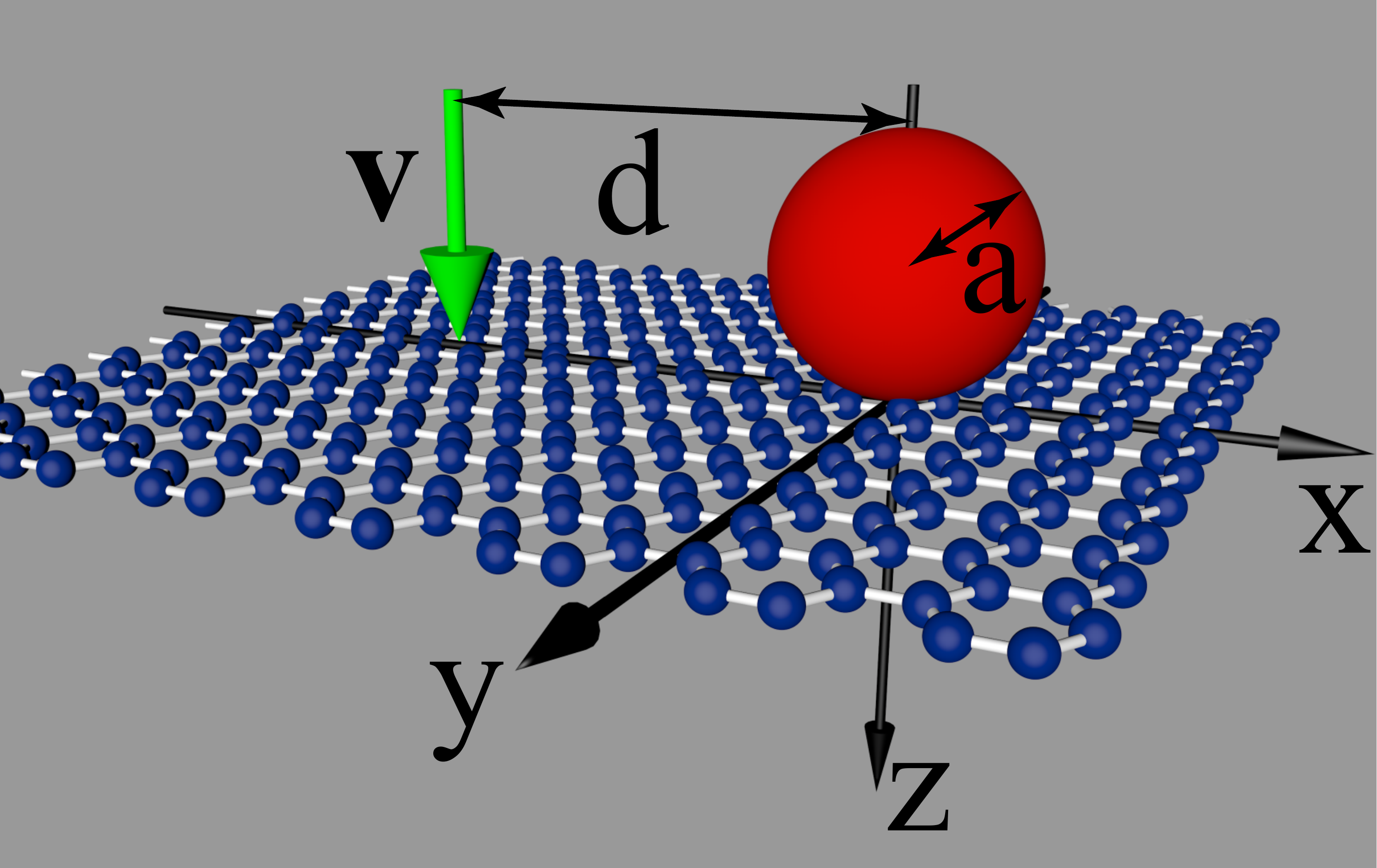}
\caption{Geometry of the electron-graphene-sphere interaction.}
\end{figure}

\section{Excitation of graphene-nanoparticle system by fast electrons}
The geometry of the graphene-nanoparticle system interacting with relativistic electrons is sketched in Fig.1. The graphene mono-layer lies on the plane $z=0$ and it is embedded in a homogeneous transparent medium whose real dielectric constant is $\varepsilon$. The plasmonic nanoparticle of radius $a$ has dielectric permittivity $\varepsilon _{\rm NP} (\omega)$ and it lies upon the graphene sheet with its center located at ${\bf r}_{\rm NP} = - a{\bf{\hat e}}_z$. An electron of velocity ${\bf{v}} = v{\bf{\hat e}}_z$ normally impinges the graphene sheet at an impact parameter $d$ with respect to the nanosphere center. In order to evaluate the radiation emitted by the system by cathodoluminescence, we first examine the electromagnetic interaction of the moving charge with the graphene monolayer and subsequently we incorporate the effect of the nanoparticle resorting to the dipole approximation. We hereafter label with $\parallel$ a vector lying in the $xy$ plane (i.e. ${\bf{A}}_\parallel = A_x {\bf{\hat e}}_x  + A_y {\bf{\hat e}}_y$) and we label with a subscript $\omega$ a frequency domain quantity by adopting the spectral analysis 
\begin{equation}
f_\omega  = \frac{1}{{2\pi }}\int {dt} \;e^{i\omega t} f\left( t \right).
\end{equation}

\subsection{Electron-graphene interaction}
An electron of charge $-e<0$ moving on the trajectory ${\bf{r}}_{\rm e} \left( t \right) =  - d {\bf{\hat e}}_x  + vt{\bf{\hat e}}_z$ is equivalent, in the frequency domain, to the charge and current densities $\rho _\omega = -\frac{e}{{2\pi v}}\delta \left( {x + d } \right)\delta \left( y \right)e^{i\frac{\omega }{v}z}$, ${\bf{J}}_\omega  = \rho _\omega  v{\bf{\hat e}}_z$
whereas graphene hosts the surface current density ${\bf{K}}_{\omega  \parallel } = \sigma (\omega) {\bf{E}}_{\omega  \parallel }$ where ${\bf{E}}_{\omega  \parallel }$ is the in-plane part of totale electric field at $z=0$. Direct solution of Maxwell equations in the frequency domain with the above charge and current densities together with the boundary conditions at $z=0$ (continuity of the electric field tangential component and discontinuity of the magnetic field tangential component produced by the graphene surface current density) leads to the electric field 
\begin{equation} \label{Field_q_GRA}
{\bf{E}}_\omega ^{\left(\rm eg\right)}  = {\bf{E}}_\omega ^{\rm \left( e \right)}  + {\bf{E}}_\omega ^{\left( \rm {g} \right)},
\end{equation}
where
\begin{eqnarray}  \label{fields}
 {\bf{E}}_\omega ^{\rm \left( e \right)}  &=& \frac{{  E_{\omega 0} }}{{\varepsilon \beta ^2 \gamma }}e^{i \frac{\omega}{v} z } \left[ { - K_1 \left( {\frac{{\omega \rho }}{{v\gamma }}} \right){\bf{\hat e}}_\rho   + \frac{i}{\gamma }K_0 \left( {\frac{{\omega \rho }}{{v\gamma }}} \right){\bf{\hat e}}_z } \right], \nonumber \\ 
 {\bf{E}}_\omega ^{\left( \rm {g} \right)}  &=& \frac{{  E_{\omega 0} }}{{\varepsilon \beta }}\int\limits_0^\infty  {dk_ \parallel  } e^{ik_z \left| z \right|} \frac{{k_ \parallel ^2 }}{{k_0 }}\frac{{k_z J_1 \left( {k_ \parallel  \rho } \right){\bf{\hat e}}_\rho   + i \: {\mathop{\rm sign}} \left( z \right)k_ \parallel  J_0 \left( {k_ \parallel  \rho } \right){\bf{\hat e}}_z }}{{\left[ {k_ \parallel ^2  + \left( {\frac{\omega }{{v\gamma }}} \right)^2 } \right]\left( {k_z  + k_0 \frac{{2\varepsilon }}{{Z_0 \sigma }}} \right)}},
\end{eqnarray}
\begin{figure}[!t] \label{Fig2}
\center
\includegraphics[width=1\textwidth]{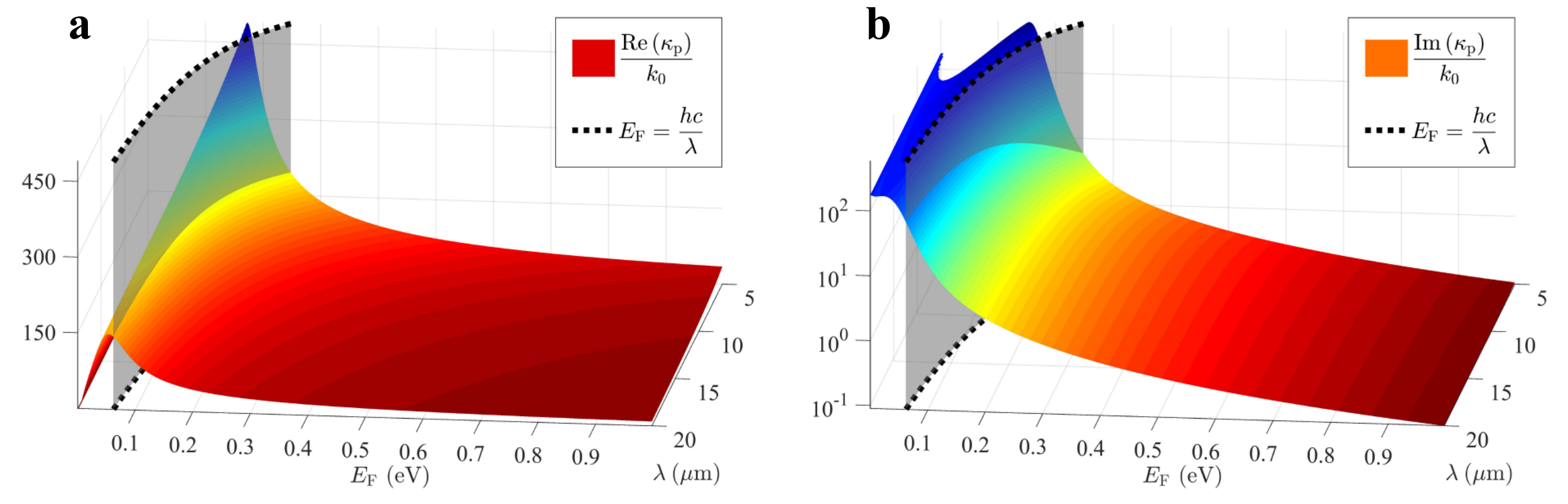}
\caption{({\bf a}) Real and ({\bf b}) imaginary part of the GPP pole $\kappa_{\rm p}$ (normalized with the vacuum wavenumber $k_0$) for $\varepsilon = 2$. On the dashed curve the Fermi energy $E_{\rm F}$ is equal to the photon energy $hc / \lambda$.}
\end{figure}
where  $\beta = \frac{v}{c}$, $\gamma  = \frac{1}{\sqrt{1 - \varepsilon \beta ^2}}$ is the Lorentz contraction factor, $Z_0  = \sqrt {\frac{\mu _0}{\varepsilon _0 }}$ is the vacuum impedance, $E_{\omega 0}  = e \frac{k_0 Z_0}{4\pi ^2}$ and $k_z  = \sqrt {k_0^2 \varepsilon  - k_ \parallel ^2 }$ with ${\mathop{\rm Im}\nolimits}  \left(k_z \right) \geq 0$. Here cylindrical coordinates coaxial with the charge trajectory have been introduced through $\rho  = \sqrt {\left( {x + d } \right)^2  + y^2 }$ and ${\bf{\hat e}}_\rho   = \nabla \rho$, while $K_n$  are the modified Bessel function of the second kind and $J_n$ are the Bessel function of the first kind. 

The field ${\bf{E}}_\omega ^{\rm \left( e \right)}$ is the well-known field produced by an electron uniformly moving in a homogenous medium with permittivity $\varepsilon$ whereas ${\bf{E}}_\omega ^{\left( \rm {g} \right)}$ is a source-free field produced by the graphene sheet (accordingly vanishing for $\sigma = 0$). Note that, due to graphene rotational invariance around the charge trajectory, the field ${\bf{E}}_\omega ^{\left( \rm {g} \right)}$ lies on the radial $\rho z$ plane as  much as ${\bf{E}}_\omega ^{\left( {\rm e} \right)}$ and the whole field ${\bf{E}}_\omega ^{\left( {\rm eg}\right)}  $ is transverse magnetic (TM). We here focus on the sub-Cherenkov regime where $v < \frac{c}{\sqrt \varepsilon}$ so that ${\mathop{\rm Im}\nolimits} \left( \gamma  \right) = 0$ and  ${\bf{E}}_\omega ^{\left( {\rm e} \right)}$ displays exponentially decaying profile (through the modified Bessel functions) and it does not provide electromagnetic radiation. On the other hand, ${\bf{E}}_\omega ^{\left( \rm {g} \right)}$ is responsible for the transition radiation (TR) associated with the graphene surface charge redistribution caused by electron crossing the mono-layer (see below). The analysis of such field is simplified by noting that it can be casted as 
\begin{equation} \label{E_gra}
{\bf{E}}_\omega ^{\left( \rm {g} \right)}  = \left( {k_0^2 \varepsilon  + \nabla \nabla  \cdot } \right) {\bf \Pi} _\omega ^{\left( \rm {gra} \right)} 
\end{equation}
where
\begin{equation} \label{Hertz_gra}
{\bf \Pi} _\omega ^{\left( \rm {g} \right)}  = E_{\omega 0} \frac{{ i \: {\rm sign}\left( z \right)}}{{\varepsilon \beta k_0 }}\int\limits_0^\infty  {dk_\parallel  } \frac{{e^{ik_z \left| z \right|} k_ \parallel  J_0 \left( {k_\parallel  \rho } \right)}}{{\left[ {k_\parallel ^2  + \left( {\frac{\omega }{{v\gamma }}} \right)^2 } \right]\left( {k_z  + k_0 \frac{{2\varepsilon }}{{Z_0 \sigma }}} \right)}}{\bf{\hat e}}_z 
\end{equation}
is the Hertz vector. 

Electron velocity $v$ affects the distribution of photon wavevectors $k_\parallel$ through the characteristic Lorentzian profile of width
\begin{equation}
\Delta k_\parallel \sim \frac{\omega} {v\gamma } = \frac{k_0} { \beta \gamma }
\end{equation}
whereas graphene yields the standard Fresnel coefficient for TM polarization whose pole at the complex wavevector 
\begin{equation} \label{GPP_pole}
\kappa_{\rm p} = k_0 \sqrt {\varepsilon  - \left( {\frac{{2\varepsilon }}{{Z_0 \sigma }}} \right)^2 } 
\end{equation}
(occurring only if ${\mathop{\rm Im}\nolimits} \left( \sigma  \right) > 0$) signals the excitation of graphene plasmon polaritons (GPPs). Real and imaginary part of $\kappa_{\rm p}$, normalized with the vacuum wavenumber $k_0$, are plotted in Fig.2{\bf a} and 2{\bf b}, respectively, for $\varepsilon = 2$. It is worth noting that if $E_{\rm F} < \frac{hc}{\lambda}$ (the region at the left side of the grey surface in Fig.2{\bf a} and 2{\bf b}), the plasmon resonance at $k_\parallel = {\rm Re} \left( \kappa_{\rm p} \right) > 150 \: k_0$ falls far outside the Lorentian distribution $\Delta k_\parallel   < 10 \: k_0$ (for $\beta > 0.1$ electrons) and it has a very low quality since ${\rm Im} \left( \kappa_{\rm p}  \right) > 10 \: k_0$. In other words, if the Fermi energy is smaller than the photon energy, GPPs are effectively not excited by the relativistic electron and consequently the graphene field ${\bf \Pi} _\omega ^{\left( \rm {g} \right)}$ is not affected by electrical gating. Therefore, since electrical tunability is among our main targets, we will focus on the regime $E_{\rm F}  > \frac{{hc}}{\lambda } = \hbar \omega$.

\begin{figure}[!t] \label{Fig3}
\center
\includegraphics[width=0.6\textwidth]{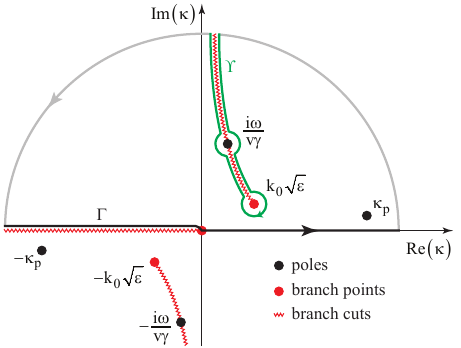}
\caption{Sommerfeld contour (black, grey and green curves) used to identify the GPP contribution to the field produced by graphene at the electron crossing.}
\end{figure}
In the chosen $E_{\rm F}  > \hbar \omega$ regime, the local model for the graphene surface conductivity of Eq.(\ref{sigma}) is fully adequate to describe the interaction with relativistic electrons. In fact, from Eq.(\ref{Hertz_gra}), the broadest photon wavevectors distribution occurs at the graphene plane $z=0$, it has a width of the order of $\Delta k_\parallel  < 10 \: k_0$ (for $\beta > 0.1$ electrons) and it hosts the additional GPP peak at $k_\parallel = {\mathop{\rm Re}\nolimits} \left( \kappa_{\rm p} \right) < 150 \: k_0$ (see Fig.2{\textbf a}) so that Eq.(\ref{locality2}) is fully satisfied.
 
The integral expression in Eq.(\ref{Hertz_gra}) is also useful to identify the GPP contribution to the graphene field which is sufficiently accurate in the near filed and far from the electron trajectory (plasmon pole approximation, see below). By using the well-known relation $J_0 \left( {k_\parallel  \rho } \right) = \frac{1}{2}\left[ {H_0^{\left( 1 \right)} \left( {k_\parallel  \rho } \right) - H_0^{\left( 1 \right)} \left( {e^{i\pi } k_\parallel  \rho } \right)} \right]$ where $H_0^{\left( 1 \right)} \left( \zeta \right)$ is the analytic continuation from the positive real axis of the Hankel function of the first kind of order $0$, Eq.(\ref{Hertz_gra}) can be casted as 
\begin{equation} \label{contour1}
\Pi _{\omega z}^{\left( {\rm g} \right)}  = E_{\omega 0} \frac{{i \: {\rm sign}\left( z \right)}}{{2\varepsilon \beta k_0 }}\int_\Gamma {d\kappa } \;\frac{{e^{ik_z \left| z \right|} \kappa H_0^{\left( 1 \right)} \left( {\kappa \rho } \right)}}{{\left[ {\kappa ^2  + \left( {\frac{\omega }{{v\gamma }}} \right)^2 } \right]\left( {k_z  + k_0 \frac{{2\varepsilon }}{{Z_0 \sigma }}} \right)}}
\end{equation}
where the contour is performed along the upper side of the real axis ($\Gamma$) due to the branch cut of $H_0^{\left( 1 \right)} \left( {\kappa \rho } \right)$ along the negative real axis (see Fig.3). Due to its asymptotic $|\kappa| \rightarrow \infty$ behavior, $H_0^{\left( 1 \right)} \left( {\kappa \rho } \right) \approx \sqrt {\frac{2}{{\pi \kappa \rho }}} \exp \left( {i \kappa \rho  - i \frac{\pi }{4}} \right)$, the Hankel function asymptotically vanishes in the upper half-plane so that, in view of the Jordan's lemma, we require ${e^{ik_z \left| z \right|} }$ to asymptotically vanish by choosing the Riemann sheet of $k_z  = \sqrt {k_0^2 \varepsilon  - \kappa ^2 }$ uniformly satisfying ${\rm Im} \left( k_z \right) \ge 0$ (i.e. $\sqrt \zeta   = \sqrt {\left| \zeta  \right|} \exp \left[ {\frac{i}{2}\arg \left( \zeta  \right)} \right]$, with $0 \le \arg \left( \zeta  \right) < 2\pi$, with branch cut at ${\rm Im} \left(\zeta\right) = 0$, ${\rm Re} \left(\zeta\right) > 0$). For mathematical convenience, we let $\varepsilon$ to have a small positive imaginary part, so that $k_z$ has branch points at $\kappa  =  \pm k_0 \sqrt \varepsilon$ close to the real axis and branch cuts along the curve ${\mathop{\rm Im}\nolimits} \left( {k_0^2 \varepsilon  - \kappa ^2 } \right) = 0$, ${\mathop{\rm Re}\nolimits} \left( {k_0^2 \varepsilon  - \kappa ^2 } \right) > 0$ comprising two  hyperbola portions asymptotically approaching the imaginary axis (see Fig.3). The integrand in Eq.(\ref{contour1}) has four simple poles, two GPP poles at $\kappa = \pm \kappa_{\rm p}$ (see Eq.(\ref{GPP_pole})) and two electronic poles at $\kappa  =  \pm \frac{{i\omega }}{{v\gamma }}$ close to the imaginary axis and lying on the branch cuts (since $ k_0^2 \varepsilon  - \left( { \pm \frac{{i\omega }}{{v\gamma }}} \right)^2  = \left( {\frac{\omega }{v}} \right)^2$ is real and positive). Residue theorem applied to the Sommerfeld contour reported in Fig.3 (black, grey and green curves), together with Jordan's lemma, implies that the integral along $\Gamma$ equals $2\pi i$ times the residue at $\kappa_{\rm p}$ minus the integral over the contour $\Upsilon$ (green curve surrounding the branch cut), so that Eq.(\ref{contour1}) yields
\begin{equation} \label{contour2}
\Pi _{\omega z}^{\left( {g} \right)}  =   E_{\omega 0} \frac{{\pi \: {\rm sign}\left( z \right)}}{{\varepsilon \beta k_0 }}\left\{ {\left[ {\frac{{  e^{ik_z \left| z \right|} k_z H_0^{\left( 1 \right)} \left( {\kappa \rho } \right)}}{{\kappa ^2  + \left( {\frac{\omega }{{v\gamma }}} \right)^2 }}} \right]_{\kappa  = \kappa _{\rm p} }  + \frac{1}{{2\pi i }}\int\limits_\Upsilon  {d\kappa } \;\frac{{e^{ik_z \left| z \right|} \kappa H_0^{\left( 1 \right)} \left( {\kappa \rho } \right)}}{{\left[ {\kappa ^2  + \left( {\frac{\omega }{{v\gamma }}} \right)^2 } \right]\left( {k_z  + k_0 \frac{{2\varepsilon }}{{Z_0 \sigma }}} \right)}}} \right\}.
\end{equation}
The first term is evidently the field of the GPP excited by the electron which is closely confined to the graphene plane with evanescent decay length $\sim \frac{1}{{{\mathop{\rm Re}\nolimits} \left( {\kappa _{\rm p} } \right)}}$ (since $k_z \left( {\kappa _{\rm p} } \right) \simeq  i\kappa _{\rm p})$) and displaying a radially oscillating asymptotic profile with period $\sim \frac{{2\pi }}{{{\mathop{\rm Re}\nolimits} \left( {\kappa _{\rm p} } \right)}}$ and decay length $\sim \frac{1}{{{\mathop{\rm Im}\nolimits} \left( {\kappa _{\rm p} } \right)}}$. The second integral term in Eq.(\ref{contour2}) is responsible in far field for the transition radiation produced by the electron crossing whereas, close to the graphene plane, it is tightly confined around the electron trajectory with the same radial decay length of the electron field ${\bf{E}}_\omega ^{\rm \left( e \right)}$. In fact, for $z=0$ and $\rho \rightarrow \infty$ the leading contribution to the integral comes from the infinitesimal circle around the pole $\frac{i \omega}{v \gamma}$ since the upper and lower portions of the contour $\Upsilon$ provide negligible contributions ($H_0^{\left( 1 \right)} \left( {\kappa \rho } \right)$ has a very fast exponential decay over the upper portion and it very rapidly oscillates over the lower portion). Hence, performing the integral over the infinitesimal circle $\kappa = \frac{i \omega}{v \gamma} + \eta e^{i \phi}$ ($\eta \rightarrow 0^+$), we get
\begin{equation}
\frac{1}{{2\pi i}}\int\limits_\Upsilon  {d\kappa } \;\frac{ {e^{ik_z \left| z \right|} \kappa H_0^{\left( 1 \right)} \left( {\kappa \rho } \right)}}{{\left[ {\kappa ^2  + \left( {\frac{\omega }{{v\gamma }}} \right)^2 } \right]\left( {k_z  + k_0 \frac{{2\varepsilon }}{{Z_0 \sigma }}} \right)}} \simeq  - \;\frac{{K_0 \left( {\frac{{\omega \rho }}{{v\gamma }}} \right)}}{{i\pi \left( {k_0 \frac{{2\varepsilon }}{{Z_0 \sigma }} + \frac{\omega }{v}} \right)}}
\end{equation}
which displays the same vanishing exponential profile of the electron field  ${\bf{E}}_\omega ^{\rm \left( e \right)}$ in the first of Eqs.(\ref{fields}). 

It is worth noting for our later purposes that, in the chosen regime of Fermi energy grater than the photon energy (where GPPs are effectively excited), the field close the graphene plane and radially far from the electron trajectory is dominated by the GPP contribution (since all the other terms display radial exponential decay). More precisely, if the condition $\frac{{\omega \rho }}{{v\gamma }} \gg 1$ is satisfied, the so called plasmon pole approximation holds and the field ${\bf{E}}_\omega ^{\left( {\rm eg}\right)}$ of Eq.(\ref{Field_q_GRA}), from Eq.(\ref{E_gra}) and the GPP term of Eq.(\ref{contour2}), reduces to
\begin{equation} 
{\bf{E}}_\omega ^{\left( {\rm eg}\right)}  = E_{\omega 0} \frac{\pi }{{\varepsilon \beta k_0 }}\left[ {e^{ik_z \left| z \right|} \kappa k_z \frac{{ - ik_z H_1^{\left( 1 \right)} \left( {\kappa \rho } \right){\bf{\hat e}}_\rho   + {\rm sign}\left( z \right)\kappa H_0^{\left( 1 \right)} \left( {\kappa \rho } \right){\bf{\hat e}}_z }}{{\kappa ^2  + \left( {\frac{\omega }{{v\gamma }}} \right)^2 }}} \right]_{\kappa  = \kappa _{\rm p} }
\end{equation}
which, using the relation $k_z \left( {\kappa _{\rm p} } \right) \simeq  i\kappa _{\rm p}$, can be casted as
\begin{equation} \label{PlaField}
{\bf{E}}_\omega ^{\left( {\rm eg} \right)}  = E_{\omega 0} \frac{{i\pi }}{{\varepsilon \beta k_0}}\frac{{\kappa _{\rm p}^3 \left[ {H_1^{\left( 1 \right)} \left( {\kappa _{\rm p} \rho } \right){\bf{\hat e}}_\rho   + \frac{z}{{\left| z \right|}}H_0^{\left( 1 \right)} \left( {\kappa _{\rm p} \rho } \right){\bf{\hat e}}_z } \right]}}{{  {\kappa _{\rm p}^2  + \left( {\frac{\omega }{{v\gamma }}} \right)^2 } }}e^{ - \kappa _{\rm p} \left| z \right|}.
\end{equation}

\subsection{Nanoparticle excitation}
We consider a plasmonic nanoparticle whose dielectric permittivity is described by the Drude model $\varepsilon _{\rm NP} (\omega) = 1 - \frac{\omega_{\rm p}^2}{\omega^2 + i \omega \Gamma}$ which accurately applies to transparent conductors with plasma frequency $\omega_{\rm p}$ in the mid-infrared. The nanoparticle-graphene evanescent coupling entails the hybridization of nanoparticle localized plasmons (NLPs) and GPPs thus yielding hybrid plasmonic modes which, in the presence of the moving electron, are excited by the field ${\bf{E}}_\omega ^{\left( {\rm eg}\right)}$ discussed in the previous section. 

Since the radius $a$ is much smaller than the mid-infrared wavelengths, we here resort to the electrostatic (no-retarded) approximation where the nanoparticle is modelled by a point dipole located at ${\bf r}_{\rm c} = -a \hat{\bf e}_z$ whose dipole moment (in the frequency domain) is ${\bf{p}}_\omega   = \alpha {\bf{E}}_\omega ^{\left( {\rm ext} \right)}$ where ${\bf{E}}_\omega ^{\left( {\rm ext} \right)}$ is the field experienced by the dipole (without self-field) and $\alpha  = 4\pi \varepsilon _0 \varepsilon  a^3 \left( {\frac{{\varepsilon _{\rm NP}  - \varepsilon }}{{\varepsilon _{\rm NP}  + 2\varepsilon }}} \right)$ is the well-known polarizability of the sphere. Due to the presence of the graphene sheet at $z=0$, the field radiated by the point dipole is
\begin{equation} \label{Field_p}
{\bf{E}}_\omega ^{\left( {\rm NP} \right)}  = \left\{ \uptheta \left( { - z} \right)  {\left[ {G^{\left( {\rm i} \right)}  + G^{\left( {\rm r} \right)} } \right]+ \uptheta \left( z \right) \left[ {G^{\left( {\rm t} \right)} } \right]} \right\}{\bf{p}}_\omega  
\end{equation}
where 
\begin{eqnarray} \label{irt}
 G ^{\left( {\rm i} \right)}  &=& \int {d^2 {\bf{k}}_\parallel  e^{i{\bf{k}}_\parallel   \cdot {\bf{r}}_\parallel  } } e^{ik_z \left| {z + a} \right|} \left[ {1 - {\mathop{\rm sign}} \left( {z + a} \right)\frac{{{\bf{\hat e}}_z {\bf{k}}_\parallel ^{\rm T} }}{{k_z }}} \right]\left[ {i\frac{{k_0^2 \varepsilon I_\parallel   - {\bf{k}}_\parallel  {\bf{k}}_\parallel ^{\rm T}  - {\mathop{\rm sign}} \left( {z + a} \right)k_z {\bf{k}}_\parallel  {\bf{\hat e}}_z^{\rm T} }}{{8\pi ^2 \varepsilon _0 \varepsilon k_z }}} \right], \nonumber  \\ 
 G^{\left( {\rm r} \right)}  &=& \int {d^2 {\bf{k}}_\parallel  e^{i{\bf{k}}_\parallel   \cdot {\bf{r}}_\parallel  } } e^{ik_z \left( { - z + a} \right)} \left( {1 + \frac{{{\bf{\hat e}}_z {\bf{k}}_\parallel ^{\rm T} }}{{k_z }}} \right)\left[ {r_{\rm TE} \left( {1 - \frac{{{\bf{k}}_\parallel  {\bf{k}}_\parallel ^{\rm T} }}{{k_\parallel ^2 }}} \right) + r_{\rm TM} \left( {\frac{{{\bf{k}}_\parallel  {\bf{k}}_\parallel ^{\rm T} }}{{k_\parallel ^2 }}} \right)} \right]\left( {i\frac{{k_0^2 \varepsilon I_\parallel   - {\bf{k}}_\parallel  {\bf{k}}_\parallel ^{\rm T}  - k_z {\bf{k}}_\parallel  {\bf{\hat e}}_z^{\rm T} }}{{8\pi ^2 \varepsilon _0 \varepsilon k_z }}} \right), \nonumber   \\ 
 G^{\left( {\rm t} \right)}  &=& \int {d^2 {\bf{k}}_\parallel  e^{i{\bf{k}}_\parallel   \cdot {\bf{r}}_\parallel  } } e^{ik_z \left( {z + a} \right)} \left( {1 - \frac{{{\bf{\hat e}}_z {\bf{k}}_\parallel ^{\rm T} }}{{k_z }}} \right)\left[ {t_{\rm TE} \left( {1 - \frac{{{\bf{k}}_\parallel  {\bf{k}}_\parallel ^{\rm T} }}{{k_\parallel ^2 }}} \right) + t_{\rm TM} \left( {\frac{{{\bf{k}}_\parallel  {\bf{k}}_\parallel ^{\rm T} }}{{k_\parallel ^2 }}} \right)} \right]\left( {i\frac{{k_0^2 \varepsilon I_\parallel   - {\bf{k}}_\parallel  {\bf{k}}_\parallel ^{\rm T}  - k_z {\bf{k}}_\parallel  {\bf{\hat e}}_z^{\rm T} }}{{8\pi ^2 \varepsilon _0 \varepsilon k_z }}} \right). \nonumber \\ 
\end{eqnarray}
Here the dyadic notation $( {{\bf{ab}}^{\rm T} } ){\bf{c}} = \left( {{\bf{b}} \cdot {\bf{c}}} \right) {\bf{a}}$ has been used,  $I_\parallel   = {\bf{\hat e}}_x {\bf{\hat e}}_x^{\rm T}  + {\bf{\hat e}}_y {\bf{\hat e}}_y^{\rm T}$ and the reflection and trasmission coeffients for TE and TM waves are
\begin{equation} 
\begin{array}{*{20}c}
   {r_{\rm TE}  =  - \frac{\displaystyle {k_0 \frac{{Z_0 \sigma }}{2}}}{\displaystyle {k_z  + k_0 \frac{{Z_0 \sigma }}{2}}},} & {t_{\rm TE}  = \frac{\displaystyle{k_z }}{\displaystyle{k_z  + k_0 \frac{{Z_0 \sigma }}{2}}},} & {r_{\rm TM}  =  - \frac{\displaystyle{k_z }}{\displaystyle{k_z  + k_0 \frac{{2\varepsilon }}{{Z_0 \sigma }}}},} & {t_{\rm TM}  = \frac{\displaystyle{k_0 \frac{{2\varepsilon }}{\displaystyle{Z_0 \sigma }}}}{\displaystyle{k_z  + k_0 \frac{{2\varepsilon }}{{Z_0 \sigma }}}}.}  \\
\end{array}
\end{equation}
The first term in Eq.(\ref{Field_p}) is the stadard dipole field $G^{\left( {\rm i} \right)} {\bf{p}}_\omega   = \left( {k_0^2 \varepsilon  + \nabla \nabla  \cdot } \right)\left( {\frac{1}{{4\pi \varepsilon _0 \varepsilon }}\frac{{e^{ik_0 \sqrt \varepsilon  \left| {{\bf{r}} + a{\bf{\hat e}}_z } \right|} }}{{\left| {{\bf{r}} + a{\bf{\hat e}}_z } \right|}}{\bf{p}}_\omega  } \right)$ in the angular spectrum representation whereas $G^{\left( {\rm r} \right)} {\bf{p}}_\omega$ and $G^{\left( {\rm t} \right)} {\bf{p}}_\omega$ are the reflected and transmitted fields, respectively, produced by the graphene sheet (and accordingly $G^{\left( {\rm r} \right)}=0$ and $G^{\left( {\rm t} \right)}=G^{\left( {\rm i} \right)}$ for $\sigma = 0$, since in this case $r_{\rm TE} = r_{\rm TM} =0$ and $t_{\rm TE} = t_{\rm TM} =1$). Evidently, the TM reflection and transmission coefficients have the plasmon pole $\kappa_{\rm p}$ (see Eq.(\ref{GPP_pole})) which signals the well-known ability of the nano-antenna to excite GPPs.

Note that, due to the factor $e^{ik_z a}$ in the second and third of Eqs.(\ref{irt}), the broadest photon wavevectors distribution at the graphene plane $z=0$ has a width of the order of $\frac{1}{a} < 105\,k_0$ (for $a = 30 \: {\rm nm}$ and $\lambda < 20 \: {\rm \mu m}$) with the same GPP peak discussed in the above setion. Therefore, in the chosen $E_{\rm F}  > \hbar \omega$ regime, Eq.(\ref{locality2}) is fully satisfied and nonlocal effects do not play any role in the nanoparticle-graphene interaction.

In the presence of the moving electron, the overall field is
\begin{equation} \label{totalfield}
{\bf{E}}_\omega   = {\bf{E}}_\omega ^{\left( {\rm eg} \right)}  + {\bf{E}}_\omega ^{\left( \rm NP \right)} 
\end{equation}
and the field experienced by the dipole is ${\bf{E}}_\omega ^{\left( {\rm ext} \right)}  = \left[ {{\bf{E}}_\omega ^{\left( {\rm eg} \right)}  + G^{\left( {\rm r} \right)} {\bf{p}}_\omega  } \right]_{{\bf{r}} = {\bf{r}}_{\rm NP} } $ so that, using the nanosphere polarizability $\alpha$, we get
\begin{equation}
{\bf{p}}_\omega   = \left[ {\frac{1}{{\frac{1}{\alpha } - G^{\left( {\rm r} \right)} }}{\bf{E}}_\omega ^{\left( {\rm eg} \right)} } \right]_{{\bf{r}} = {\bf{r}}_{\rm NP} } 
\end{equation} 
for the induced dipole moment. Since the field ${{\bf{E}}_\omega ^{\left( {\rm eg} \right)} }$ lies on the radial $\rho z$ plane, this equation implies that the dipole moment has only $x$- and $z$- components (i.e. ${\bf{p}}_\omega   = p_{\omega x} {\bf{\hat e}}_x  + p_{\omega z} {\bf{\hat e}}_z$) given by
\begin{eqnarray} \label{momentcomp}
 p_{\omega x}  &=& \frac{\displaystyle {E_{\omega \rho }^{\left( {\rm eg} \right)} \left( {{\bf{r}}_{\rm NP} } \right)}}{\displaystyle {\frac{1}{\alpha } - \frac{i}{{8\pi \varepsilon _0 \varepsilon }}\int\limits_0^\infty  {dk_\parallel  } e^{i2k_z a} \left( {r_{\rm TE} \frac{{k_\parallel  k_0^2 \varepsilon }}{{k_z }}\, + r_{\rm TM} k_\parallel  k_z } \right)}}, \nonumber \\ 
 p_{\omega z}  &=& \frac{\displaystyle{E_{\omega z}^{\left( {\rm eg} \right)} \left( {{\bf{r}}_{\rm NP}  } \right)}}{\displaystyle{\frac{1}{\alpha } - \frac{i}{{8\pi \varepsilon _0 \varepsilon }}\int\limits_0^\infty  {dk_\parallel  } e^{i2k_z a} \left(- {r_{\rm TM} \frac{{2k_\parallel ^3 }}{{k_z }}} \right)}}
 \end{eqnarray}
where angular integration have been performed in $G^{\left( \rm r \right)}$. Equation (\ref{totalfield}), with the help of Eqs.(\ref{momentcomp}), fully describe the field accompanying the interaction of the relativistic electron with the graphene-nanoparticle system. Hybrid plasmonic resonances of the nanoparticle-graphene system are identified by the poles of ${\bf{p}}_\omega$ so that Equations (\ref{momentcomp}) reveal that the fast electron is able to excite two different hybrid plasmonic modes whose dipole moments are purely $x$- and $z$- polarized, respectively. In order for the denominators of Eqs.(\ref{momentcomp}) to be very small, the $\frac{1}{\alpha}$ and the integral contributions have to be comparable which requires the both NLPs and GPPs have to be excited. Therefore the hybrid plasmonic resonances appear spectrally close to the nanoparticle resonance wavelength in the Fermi energy range where graphene plasmonic resonance occurs.

The overall field of Eq.(\ref{totalfield}) turns out to be highly sensible to the graphene Fermi energy (electric tunability) for two main reasons. First the GPP peak appears in the wavevector spectral distributions of all the graphene reaction fields, i.e. the one directly induced by the electron ${\bf{E}}_\omega ^{\left( {\rm g} \right)}$ (second of Eqs.(\ref{fields})) and the two fields produced by the dipole $G^{\left( {\rm r} \right)}{\bf{p}}_\omega$ and $G^{\left( {\rm t} \right)}{\bf{p}}_\omega$ (Eq.(\ref{Field_p})). Second, and most importantly for our purposes, the dipole field ${\bf{E}}_\omega ^{\left( {\rm NP} \right)}$ directly experiences the above discussed hybrid plasmonic resonances, with a particularly spectacular impact, since they are carried by 
${\bf{p}}_\omega$ thus uniformly enhancing the overall wavevector spectral distribution of ${\bf{E}}_\omega ^{\left( {\rm NP} \right)}$.

\section{Directionality of cathodoluminescence emission and its tuning}
The field ${\bf{E}}_\omega$ of Eq.(\ref{totalfield}) has spectral components with $k_\parallel < k_0 \sqrt{\varepsilon}$ which survive in the far field. This physically corresponds to emission of radiation by the target (here the graphene-nanoparticle system) upon interaction with the fast electron, a well-known fact usually referred to as cathodoluminescence (CL). We here investigate the tunability of the spectral CL emission, provided by the graphene Fermi energy, with emphasis on the angular distribution of the radiation pattern.

\subsection{Far field and spectral-angular distribution of the CL emission}
Since we are considering the sub-Cherenkov regime, the electron field ${\bf E}^{\left(\rm e\right)}_{\omega}$ (in the first of Eqs.(\ref{fields})) does not contribute to the emitted radiation so that, after suppressing it, the total field of Eq.(\ref{totalfield}) in the far field ($k_0 r \rightarrow \infty$) reduces to
\begin{equation} \label{farfield}
{\bf{E}}_\omega   = \frac{{e^{i\sqrt \varepsilon \left(k_0 r\right)} }}{{k_0 r}}\left[ {{\bf{f}}^{\left( {\rm g} \right)}  + {\bf{f}}^{\left( {\rm NP} \right)} } \right]
\end{equation}
where
\begin{eqnarray} \label{amplitudes}
 {\bf{f}}^{\left( {\rm g} \right)}  &=& E_{\omega 0} e^{i\sqrt \varepsilon  \left( {k_0 d} \right)\sin \theta \cos \varphi } \frac{{\left( {\frac{{Z_0 \sigma }}{{2\sqrt \varepsilon  }}} \right)}}{{1 + \left( {\frac{{Z_0 \sigma }}{{2\sqrt \varepsilon  }}} \right)\left| {\cos \theta } \right|}}\left( {\frac{{\beta \sin \theta \cos \theta }}{{\varepsilon \beta ^2 \cos ^2 \theta  - 1}}} \right){\bf{\hat e}}_\theta  , \nonumber \\ 
 {\bf{f}}^{\left( {\rm NP} \right)}  &=& \uptheta \left( { - \cos \theta } \right)  e^{i\sqrt \varepsilon  \left( {k_0 a} \right)\cos \theta } \left( {{\bf{\hat e}}_\theta  {\bf{\hat e}}_\theta ^{\rm T}  + {\bf{\hat e}}_\varphi  {\bf{\hat e}}_\varphi ^{\rm T} } \right) \frac{{k_0^3 {\bf{p}}_\omega}}{{4\pi \varepsilon _0 }}     + \nonumber \\ 
  &+& \uptheta \left( { - \cos \theta } \right) e^{ - i\sqrt \varepsilon  \left( {k_0 a} \right)\cos \theta } \left[ {\frac{{\left( {\frac{{Z_0 \sigma }}{{2\sqrt \varepsilon  }}} \right)\cos \theta }}{{1 - \left( {\frac{{Z_0 \sigma }}{{2\sqrt \varepsilon  }}} \right)\cos \theta }}\left( {\cos 2\theta \: {\bf{\hat e}}_\theta  {\bf{\hat e}}_\theta ^{\rm T}  + \sin 2\theta \: {\bf{\hat e}}_\theta  {\bf{\hat e}}_r^{\rm T} } \right) + \frac{{\left( {\frac{{Z_0 \sigma }}{{2\sqrt \varepsilon  }}} \right)}}{{\cos \theta  - \left( {\frac{{Z_0 \sigma }}{{2\sqrt \varepsilon  }}} \right)}}{\bf{\hat e}}_\varphi  {\bf{\hat e}}_\varphi ^{\rm T} } \right] \frac{{k_0^3 {\bf{p}}_\omega}}{{4\pi \varepsilon _0 }}  +  \nonumber \\ 
  &+& \uptheta \left( {\cos \theta } \right) e^{i\sqrt \varepsilon  \left( {k_0 a} \right)\cos \theta }  \left[ {\frac{1}{{1 + \left( {\frac{{Z_0 \sigma }}{{2\sqrt \varepsilon  }}} \right)\cos \theta }}{\bf{\hat e}}_\theta  {\bf{\hat e}}_\theta ^{\rm T}  + \frac{{\cos \theta }}{{\cos \theta  + \left( {\frac{{Z_0 \sigma }}{{2\sqrt \varepsilon  }}} \right)}}{\bf{\hat e}}_\varphi  {\bf{\hat e}}_\varphi ^{\rm T} } \right]
\frac{{k_0^3 {\bf{p}}_\omega}}{{4\pi \varepsilon _0 }}     , 
\end{eqnarray}
in which polar spherical coordinates $\left( r,\theta,\varphi \right)$ have been introduced together with their coordinate unit vectors ${\bf{\hat e}}_r$, ${\bf{\hat e}}_\theta$ and ${\bf{\hat e}}_\varphi$. Here ${\bf{f}}^{\left( {\rm g} \right)}$ is the far field amplitude of the graphene field ${\bf E}^{\left(\rm g\right)}_{\omega}$ and it describes the transition radiation (TR) which is generated by the electron crossing the graphene sheet. Note that ${\bf{f}}^{\left( {\rm g} \right)}$ is along the ${\bf{\hat e}}_\theta$ direction, it has a phase factor accounting for the electron impact parameter $d$ and it displays a Fresnel-like coefficient (proportional to $\sigma$) modulated by standard $\beta$-dependent factor (not diverging in the sub-Cherenkov regime $\sqrt \varepsilon  \beta < 1$ we are considering). On the other hand ${\bf{f}}^{\left( {\rm NP} \right)}$ is the far-field amplitude of the dipole field ${\bf E}^{\left(\rm NP \right)}_{\omega}$ and it describes the diffraction radiation (DR) which is outcoupled from the nanoparticle excited by the field ${\bf E}^{\left(\rm eg\right)}_{\omega}$. The amplitude ${\bf{f}}^{\left( {\rm NP} \right)}$  has three contributions arising from the fields $G ^{\left( {\rm i} \right)} {\bf p}_\omega$, $G ^{\left( {\rm r} \right)} {\bf p}_\omega$ and $G ^{\left( {\rm t} \right)} {\bf p}_\omega$, respectively and it  has components both along ${\bf{\hat e}}_\theta$ and  ${\bf{\hat e}}_\varphi$ which are suitable projections of the dipole moment ${\bf p}_\omega$.

The total energy emitted by CL per incoming electron is $U = \int\limits_{ - \infty }^\infty  {dt} \int {d\Omega } \: r^2 {\bf{\hat e}}_r  \cdot \left[ {{\bf{E}}\left( {{\bf{r}},t} \right) \times {\bf{H}}\left( {{\bf{r}},t} \right)} \right]$ which, resorting to the frequency domain, can be suitably casted as a superposition of photon energy quanta $\frac{hc}{\lambda}$, i.e. 
\begin{equation}
U = \int\limits_0^\infty  {d\lambda } \left( {\frac{{hc}}{\lambda }} \right)\int {d\Omega } \;\frac{{dN}}{{d\Omega  d\lambda }}
\end{equation}
where $\frac{{dN}}{{ d\Omega d\lambda}} = \frac{{4\pi }}{{\hbar \lambda }} r^2 {\bf{\hat e}}_r  \cdot {\mathop{\rm Re}\nolimits} \left( {{\bf{E}}_\omega   \times {\bf{H}}_\omega ^* } \right)$ is the number of photons emitted per incoming electron, per per unit of solid angle of emission and per unit of photon wavelength. By using Eq.(\ref{farfield}) and the far field relation ${\bf{H}}_\omega   = \frac{{\sqrt \varepsilon  }}{{Z_0 }}{\bf{\hat e}}_r  \times {\bf{E}}_\omega$, we get the spectral-angular distribution of the photon emission probability 
\begin{equation} \label{emission}
\frac{{dN}}{{d\Omega d\lambda }} = \frac{{\lambda \sqrt \varepsilon  }}{{\pi \hbar Z_0 }}\left[ {\left| {f_\theta ^{\left( {\rm g} \right)}  + f_\theta ^{\left( {\rm NP} \right)} } \right|^2  + \left| {f_\varphi ^{\left( {\rm NP} \right)} } \right|^2 } \right]
\end{equation}
revealing that the $\theta$- components of the graphene and dipole fields interfere in the CL radiation pattern. Since TR and DR have different spatial symmetry properties, their interferece in Eq.(\ref{emission}) provides peculiar directionality traits to the overall CL emission. In addition, since the nanoparticle excitation strongly depends, at each wavelength,  on the graphene Fermi energy, it turns out that the CL emission directionality can effectively be tuned by electrical gating.

\subsection{CL emission directionality}
In order to investigate CL emission directionality more closely, we note that in our nanophotonic setup the inequalities $k_0 d \ll 1$, $k_0 a \ll 1$ and $\left| {\frac{{Z_0 \sigma }}{{2\sqrt \varepsilon  }}} \right| \ll 1$ hold in the chosen infrared range so that Eqs.(\ref{amplitudes}) reduce to
\begin{eqnarray} 
 {\bf{f}}^{\left( {\rm g} \right)}  &=& E_{\omega 0} \left( {\frac{{Z_0 \sigma }}{{2\sqrt \varepsilon  }}} \right)\left( {\frac{{\beta \sin \theta \cos \theta }}{{\varepsilon \beta ^2 \cos ^2 \theta  - 1}}} \right){\bf{\hat e}}_\theta  , \nonumber \\ 
 {\bf{f}}^{\left( {\rm NP} \right)}  &=& \left( {{\bf{\hat e}}_\theta  {\bf{\hat e}}_\theta ^{\rm T}  + {\bf{\hat e}}_\varphi  {\bf{\hat e}}_\varphi ^{\rm T} } \right)\frac{{k_0^3 {\bf{p}}_\omega  }}{{4\pi \varepsilon _0 }}.
\end{eqnarray}
Since the amplitude $f_\theta ^{\left( {\rm g} \right)}$ does not depend on $\varphi$, the TR angular distribution ($ \sim \left| {f^{\left( {\rm g} \right)} } \right|^2$) is axially symmetric around the electron trajectory with its characteristic double cone shape (see Fig.1b of the main text) of aperture $\theta_{\rm max}$ (with $\tan \theta _{\rm max }  = \sqrt {1 - \beta ^2 \varepsilon }$) and maximum $\left| {f_\theta ^{\left( {\rm g} \right)} } \right|_{\rm max }  \simeq E_{\omega 0} \frac{\beta }{2}\left( {\frac{{Z_0 \sigma }}{{2\sqrt \varepsilon  }}} \right)$. The amplitude ${\bf{f}}^{\left( {\rm NP} \right)}$ is the standard dipole far field amplitude (see Fig.1b of the main text)  and the maximum of its $\theta$-component is $\left| {f_\theta ^{\left( {\rm NP} \right)} } \right|_{\max }  \simeq \frac{{k_0^3 \left| {{\bf{p}}_\omega  } \right|}}{{4\pi \varepsilon _0 }}$. Therefore, the relative impact of TR and DR to their interference is basically measured by the ratio
\begin{equation}
R = \frac{{\left| {f_\theta ^{\left( NP \right)} } \right|_{\max } }}{{\left| {f_\theta ^{\left( g \right)} } \right|_{\max } }} \cong \frac{{4\sqrt \varepsilon  }}{{\beta Z_0 \sigma }}\frac{{k_0^3 \left| {{\bf{p}}_\omega  } \right|}}{{4\pi \varepsilon _0 E_{\omega 0} }}
\end{equation}
which can be adjusted to be close to 1 by adjusting the electron velocity. Once the condition $R \simeq 1$ is achieved, TR and DR intereference is effective and the directionality of the CL angular distribution basically stems from the relations
\begin{eqnarray} \label{relations}
f_\theta ^{\left( {\rm g} \right)} \left( {\theta } \right)
 &=&  -  f_\theta ^{\left( {\rm g} \right)} \left( {\pi  - \theta  } \right), \nonumber  \\ 
 f_\theta ^{\left( {\rm NP} \right)} \left( {\theta ,\varphi } \right) &=& \frac{{k_0^3 }}{{4\pi \varepsilon _0 }}\left( {\cos \theta \cos \varphi \: p_{\omega x}  - \sin \theta \: p_{\omega z} } \right). 
\end{eqnarray}
The first of these equations states that the TR amplitude has opposite signs in the half spaces $z>0$ and $z<0$ whereas the second equation shows that the DR amplitude does not have this property if $p_{\omega z} \neq 0$, resulting in different interference patterns in the two half-spaces. In addition the dependence of $f_\theta ^{\left( {\rm NP} \right)}$ of $\varphi$ implies that the interference is not axially symmetric around the $z$-axis with the maximum emission direction angle $\varphi_{max}$ dependent on the dipole moment components $p_{\omega x}$ and $p_{\omega z}$.

\subsection{Impact of the GPP phase on the CL directionality}
As discussed above, the tunability of the maximal CL emission direction is an interferometric effect relying on the dependence of the excitated dipole moment ${\bf p}_{\omega}$ on the Fermi energy at each wavelength. As a consequence the effect is particularly spectacular close the hybrid nanoparticle-graphene resonances where $|{\bf p}_{\omega}|$ is highly sensible to variations of $E_{\rm F}$. There is however a specific situation where the phases $\arg p_{\omega x}$ and $\arg p_{\omega z}$ play a significant role leading to an even more spectacular angular emission phenomenology. This happens when the nanoparticle is far enough from the electron trajectory that it experiences only the GPP field  whose phase is a rapidly-varying function of both $\lambda$ and $E_{\rm F}$.

To discuss this effect, we choose the impact parameter $d$ in such a way that
\begin{equation}
d > \frac{{\beta \gamma }}{{2\pi }}\lambda 
\end{equation}
for each wavelength in the considered infrared spectral domain, so that the plasmon pole approximation of Eq.(\ref{PlaField}) holds. Since in this regime $
\left| {\kappa _{\rm p} d} \right|$ is very large, we can resort to the Hankel function asymptotic behavior $H_n^{\left( 1 \right)} \left( \zeta  \right) \simeq \sqrt {\frac{2}{{\pi z}}} e^{i\left( {\zeta  - \frac{\pi }{4} - n\frac{\pi }{2}} \right)}$ (for $|\zeta| \rightarrow \infty$) so that the field at the nanoparticle (and triggering its dipole moment, see Eq.(\ref{momentcomp})) from Eq.(\ref{PlaField}) can be casted as
\begin{equation}
{\bf{E}}_\omega ^{\left( {\rm eg} \right)} \left( {{\bf{r}}_{NP} } \right) = E_{\omega 0} \sqrt {\frac{{2\pi }}{{\kappa _{\rm p} d}}} \frac{{\kappa _{\rm p}^3 e^{ - \kappa _{\rm p} a} }}{{\varepsilon \beta k_0 \left[ {\kappa _{\rm p}^2  + \left( {\frac{\omega }{{v\gamma }}} \right)^2 } \right]}}e^{i\left( {\kappa _{\rm p} d - \frac{\pi }{4}} \right)} \left( {{\bf{\hat e}}_x  + i{\bf{\hat e}}_z } \right)
\end{equation}
where the relation $k_z  \simeq i\kappa _p$ (correct in this regime) has been exploited. As expected, the GPP field turns out to be circularly polarized in the $xz$ plane (i.e. carrying transverse momentum-locked spin) and exhibiting the plasmon phase factor $e^{i\kappa _{\rm p} d }$ which is a rapidly-varying function of both $\lambda$ and $E_{\rm F}$ since ${\mathop{\rm Re}\nolimits} \left( {\kappa _{\rm p} d} \right)$ is large (see Fig.2a). Due to Eqs.(\ref{momentcomp}), both $p_{\omega x}$ and $p_{\omega z}$ turn out to be proportional to $e^{i\kappa _{\rm p} d }$ which are the only rapidly-varying phase factor. Now, from Eq.(\ref{emission}) (and the second of Eqs.(\ref{relations})) the interference term in the angular emission pattern is
\begin{equation}
2{\mathop{\rm Re}\nolimits} \left( {f_\theta ^{\left( {\rm g} \right)*} f_\theta ^{\left( {\rm NP} \right)} } \right) = \frac{{k_0^3 }}{{2\pi \varepsilon _0 }}{\mathop{\rm Re}\nolimits} \left[ {f_\theta ^{\left( g \right)*} \left( {\cos \theta \cos \varphi \;p_{\omega x}  - \sin \theta \;p_{\omega z} } \right)} \right]
\end{equation}
which, due to the plasmon phase factor $e^{i\kappa _{\rm p} d }$ in both $p_{\omega x}$ and $p_{\omega z}$, is evidently a rapidly varying function of both $\lambda$ and $E_{\rm F}$. We conclude that in the plasmon pole approximation, the CL angular distribution is highly sensible to the Fermi energy thus providing the effective ability to tune the maximal emission direction very easily through small variation of $E_{\rm F}$. Conversely, in a spectroscopic perspective, the phenomenon can also be exploited to extract the GPP phase by comparing the CL angular distribution at different Fermi energies, at each wavelength.

\section{Supporting Movies}

Dependence on the Fermi energy of various quantities related to the CLR emission considered in the main text at the on-resonance wavelength $\lambda = 6 \, {\mu m}$ (Supporting Movie S1) and the two on-resonance wavelengths $\lambda = 11.2 \, {\mu m}$ (Supporting Movie S2) and $\lambda = 12.6 \, {\mu m}$ (Supporting Movie S3). The plotted quantities evolving with the increasing Fermi energy are: 
the moduli and phases of the (normalized) NP dipole moment components $\tilde p_{\omega x}$ and $\tilde p_{\omega z}$ (extracted from Fig.2c of the main text), the angle-resolved TR emission pattern coaxially superimposed to the electron trajectory, the angle-resolved DR emission pattern, the dipole polarization ellipse ${\bf{p}}\left( t \right) = {\mathop{\rm Re}\nolimits} \left( {{\bf{p}}_\omega  e^{ - i\omega t} } \right)$ (arbitrarly rescaled for visualization purposes) superimposed to the DR pattern, the angle-resolved CLR emission pattern.